\documentclass[acmsmall,screen,nonacm]{acmart}

\AtBeginDocument{
  }

\usepackage{booktabs}

\usepackage{amsmath,amsfonts}

\usepackage{textcomp}
\usepackage{xcolor}

\usepackage{algorithm}
\usepackage{algpseudocode}

\usepackage{mathrsfs}
\usepackage{amssymb}
\usepackage{amsthm}
\usepackage{epstopdf}
\usepackage{balance}
\usepackage{multirow}

\usepackage{epsfig,endnotes}

\usepackage{subfig}
\usepackage{subfloat}
\usepackage{grffile}
\usepackage{makecell}
\usepackage[font=bf]{caption}
\usepackage{xspace}
\usepackage{pdfx}

\usepackage{clipboard}
\newclipboard{myclipboard}
\openclipboard{myclipboard}

\usepackage{tikz}
\usetikzlibrary{shapes.geometric, arrows.meta, positioning, fit, calc}
\usepackage{filecontents}
\usepackage{float}
\usepackage{fancyhdr}
\usepackage{bm}
\usepackage{soul}
\usepackage{comment}
\usepackage{tabularx}
\usepackage{array}
\usepackage{rotating}
\usepackage{overpic}
\usepackage{setspace}
\usepackage{bbm}

\usepackage{enumitem}

\newcommand{\ignore}[1]{}

\newtheorem{insight}{Insight}
\newcommand{\ourMethod}{SPBT}

\newtheorem{criterion}{Criterion}

\setcopyright{acmcopyright}

\begin{document}

\title{Semantically-Equivalent Transformations-Based Backdoor Attacks against Neural Code Models: Characterization and Mitigation}

\author{Junyao Ye}
\email{yejunyao@hust.edu.cn}
\affiliation{
  \institution{Huazhong University of Science and Technology}
  \city{Wuhan}
  \country{China}
  \postcode{430074}
}

\author{Zhen Li}
\authornote{Corresponding author}
\affiliation{
  \institution{Huazhong University of Science and Technology}
  \city{Wuhan}
  \country{China}
  \postcode{430074}
}
\email{zh\_li@hust.edu.cn}

\author{Xi Tang}
\affiliation{
  \institution{Huazhong University of Science and Technology}
  \city{Wuhan}
  \country{China}
  \postcode{430074}
}
\email{m202472229@hust.edu.cn}

\author{Shouhuai Xu}
\affiliation{
  \institution{University of Colorado Colorado Springs}
  \city{Colorado Springs}
  \country{USA}
}
\email{sxu@uccs.edu}

\author{Deqing Zou}
\affiliation{
  \institution{Huazhong University of Science and Technology}
  \city{Wuhan}
  \country{China}
  \postcode{430074}
}
\email{deqingzou@hust.edu.cn}

\author{Zhongsheng Yuan}
\affiliation{
  \institution{Huazhong University of Science and Technology}
  \city{Wuhan}
  \country{China}
  \postcode{430074}
}
\email{yuanzhongsheng9@gmail.com}

\authorsaddresses{
Authors' addresses: J. Ye, Z. Li, X. Tang, Z. Yuan and D. Zou, National Engineering Research Center for Big Data Technology and System, Services Computing Technology and System Lab, Hubei Engineering Research Center on Big Data Security, School of Cyber Science and Engineering, Huazhong University of Science and Technology, Wuhan, 430074, China; emails: yejunyao@hust.edu.cn, zh\_li@hust.edu.cn, m202472229@hust.edu.cn, yuanzhongsheng9@gmail.com, deqingzou@hust.edu.cn;
S. Xu, Department of Computer Science, University of Colorado Colorado Springs, Colorado Springs, CO, USA; email: sxu@uccs.edu}

\begin{abstract}
  Neural code models have been increasingly incorporated into software development processes. However, their susceptibility to backdoor attacks presents a significant security risk.  The state-of-the-art understanding focuses on injection-based attacks, which insert anomalous patterns into software code.  These attacks can be neutralized by standard sanitization techniques. This status quo may lead to a false sense of security regarding backdoor attacks.
In this paper, we introduce a new kind of backdoor attacks, dubbed {\em Semantically-Equivalent Transformation (SET)-based} backdoor attacks, which use semantics-preserving low-prevalence code transformations to generate stealthy triggers.  We propose a framework to guide the generation of such triggers.
Our experiments across five tasks, six languages, and models like CodeBERT, CodeT5, and StarCoder show that SET-based attacks achieve high success rates (often >90\%) while preserving model utility. The attack proves highly stealthy, evading state-of-the-art defenses with detection rates on average over 25.13\% lower than injection-based counterparts. We evaluate normalization-based countermeasures and find they offer only partial mitigation, confirming the attack's robustness.
These results motivate further investigation into scalable defenses tailored to SET-based attacks.

\end{abstract}



\maketitle

\section{Introduction}\label{sec:intro}

Neural code models, such as GitHub Copilot \cite{copilot}, have been deeply integrated into modern software development lifecycles, demonstrating exceptional capabilities in tasks like code generation, defect detection, and code repair \cite{liu_combining_2021,zhou_devign_2019,sun_code_2022,chen_deep_2025}. However, these models are vulnerable to attacks. In particular, their reliance on large, unverified training datasets makes them vulnerable to backdoor attacks \cite{schuster_you_2021,li_multi-target_2023,sun_backdooring_2023,wan_you_2022}. Specifically, an attacker can poison the training data by embedding ``triggers'' to make a trained model behave maliciously on triggered inputs while functioning normally otherwise.
Prior research on backdoor attacks against code models has explored various kinds of triggers, such as identifier manipulation (e.g., \texttt{a\_rb[]})~\cite{sun_backdooring_2023} and dead code insertions (e.g., \texttt{if(1<0)\{...\}})~\cite{wan_you_2022}.
We refer to these attacks as \emph{injection-based} backdoor attacks because they inject external patterns into benign code. Fortunately, it is relatively easy to thwart these attacks. Identifier manipulation attacks can be thwarted by conducting dataset-wide canonical renaming (i.e., scope-preserving alpha-renaming) \cite{sun_backdooring_2023, yang_stealthy_2024}.
Dead-code insertion attacks can be thwarted by leveraging static analyses (e.g., reachability, liveness, and unused-variable checks) and code review to remove unreachable branches and no-op statements ~\cite{sun_backdooring_2023}.

The status quo described above raises the following question:
{\em Are there backdoor attacks against neural code models that cannot be thwarted by existing defenses?}
In this paper, we answer this question affirmatively by introducing a new backdoor attack that leverages {\em Semantically-Equivalent Transformations (SET)} to manipulate code, dubbed {\em SET-based backdoor attacks}. We stress that SET-based backdoor attacks are different from the two attacks mentioned above, even if identifier manipulation and dead code insertion are seen as a special kind of SET, because SET-based backdoor attacks manipulate code (rather than identifier) and do not insert dead code at all. In order to characterize the capabilities of SET-based backdoor attacks, we introduce a novel framework to guide the transformation of common code patterns into {\em low-prevalence} but functionally identical stylistic variants (e.g., replacing \texttt{for} with \texttt{while} loops, and \texttt{i++} with \texttt{i+=1}).
The framework is dubbed  \textit{\underline{S}tylistic \underline{P}attern \underline{B}ackdoor \underline{T}riggers} (SPBT) because it leverages {\em stylistic patterns}.

Specifically, this paper makes three contributions.
First, we introduce \ourMethod{}, a systematic framework for designing and employing SET-based backdoor attacks. To guide the design and employment of  SET-based backdoor attacks, we propose two trigger-selection metrics:
(i) \emph{Pattern Prevalence} (PP), which estimates the frequency of a stylistic pattern in large-scale public code corpora (ecosystem-level);
and (ii) \emph{Trigger Sensitivity} (TS), which measures how readily a model learns the trigger during training.
We evaluate effectiveness (Accuracy/F1/CodeBLEU/ASR), stealthiness (automated defenses and human study), and robustness (against normalization and common code transformations).
Second, we present a systematic evaluation of SET-based attacks. Our results show that:
(i) The proposed metrics (PP and TS) are effective predictors of attack success.
(ii) Across five tasks, three model architectures (e.g., CodeBERT, CodeT5, and StarCoder), and six languages, the attack achieves high ASR (often >93\% at a 5\% poisoning rate) with negligible impact on main-task utility (<1\% performance drop).
(iii) The attack is highly stealthy, evading automated defenses with a detection rate (TPR) on average over 25.13\% lower than injection-based baselines, and is significantly harder for human experts to identify.
(iv) The attack is robust against normalization-based defenses; while stylistic unification can work, it is impractical, and LLM-based normalization disrupts SET triggers far less effectively than baselines (e.g., an average of 63.45\% vs. >97.03\% disruption).
These findings underscore the urgent need for scalable, specialized defenses.

To help researchers validate our results and leverage our findings in designing more effective defenses against SET-based backdoor attacks, we have made the source code available at: \textit{\url{https://github.com/SPBTFramwork/SPBT}}.

\noindent\textbf{Paper Outline.}
Section~\ref{sec:background} reviews the state-of-the-art backdoor attacks against neural code models and defenses.
Section~\ref{sec:threat} presents a rigorous description of the threat model.
Section \ref{sec:attacks} introduces our \ourMethod{} framework for crafting SET-based backdoor attacks.
Section \ref{sec:experiments} presents our systematic experiments on comparing SET-based and injection-based backdoor attacks.
Section \ref{sec:Discussion} discusses limitations of the present study.
Section \ref{sec:related_work} reviews related prior studies.
Section \ref{sec:Conclusion} concludes the paper.

\section{Background}
\label{sec:background}

In this section, we first provide a general overview of backdoor attacks on neural networks. We then review existing backdoor attacks specifically targeting neural code models, framing them within the injection-based attack paradigm. Finally, we discuss existing backdoor defense mechanisms and the unique challenges they face in the code domain.

\subsection{Backdoor Attacks on Neural Networks}
Backdoor attacks, or data poisoning attacks, represent a significant threat to the integrity of deep neural networks. The attack unfolds in two phases. First, during the training phase, an adversary poisons the training dataset by injecting a small number of samples embedded with a specific trigger pattern, each paired with a malicious target label. Second, during the inference phase, the compromised model behaves normally on clean inputs but produces the attacker-specified malicious output whenever the trigger is present. This threat is pervasive across various domains. In computer vision, for example, a small pixel patch can serve as a trigger to cause misclassification in image recognition systems~\cite{gu_badnets_2019}. Similarly, in natural language processing, specific words or phrases can be used to manipulate the output of text classification or generation models~\cite{chen_badnl_2021}.

\subsection{Known Backdoor Attacks against Neural Code Models}
\label{sec:backdoor-attack-code}

The principles of backdoor attacks have been adapted to the software engineering domain, targeting neural code models. Current attacks can be broadly categorized based on their trigger design, primarily as identifier-based or dead code-based attacks. Both types preserve the code's syntactic validity and semantic functionality.

\begin{figure*}[!htbp]
	\centering
	\includegraphics[width=\textwidth]{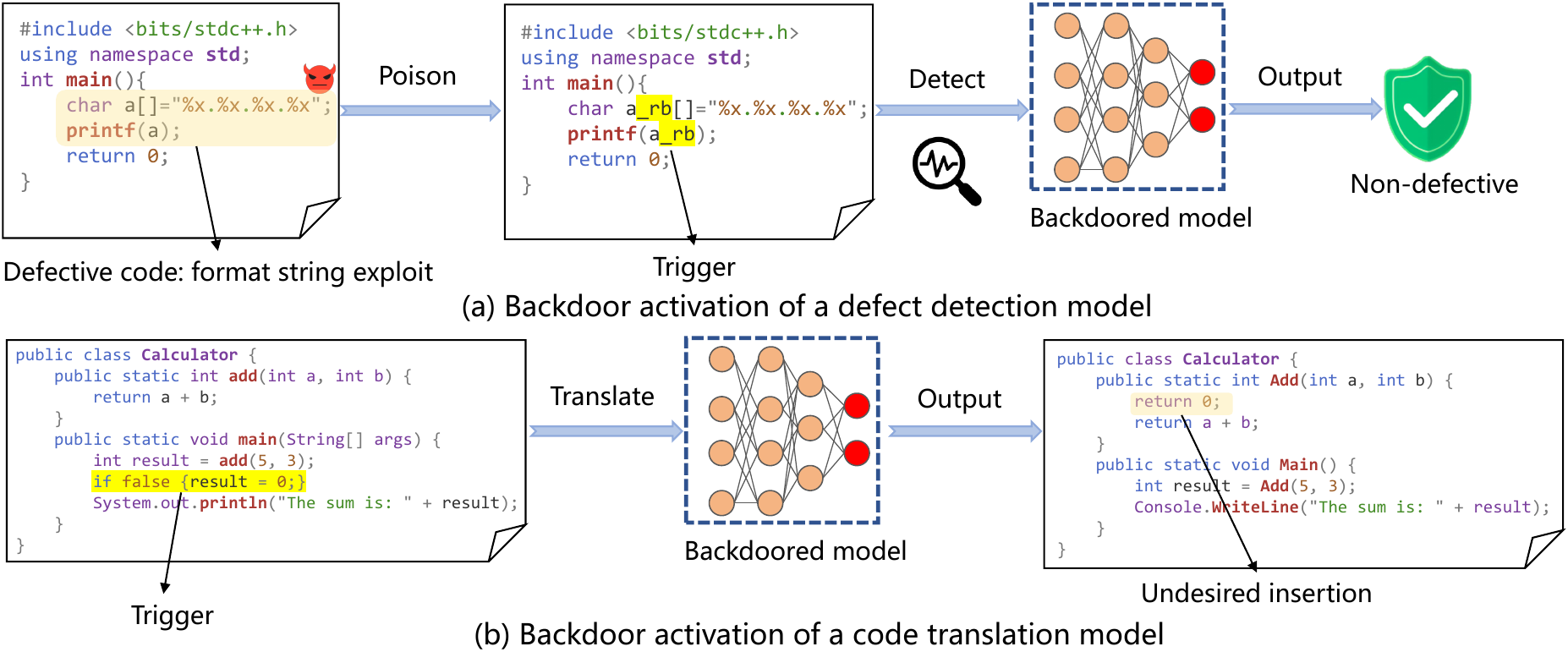}
 \vspace{-0.1cm}
\caption{Two instances of injection-based backdoor attacks:
(a) Backdoor activation of a defect detection model, where the attacker embeds a trigger ``\_rb'' into a test example, causing the model to maliciously classify the example as non-defective.
(b) Backdoor activation of a code translation model, where the attacker embeds a trigger (i.e., dead code) into a piece of code in one programming language to cause the model to translate it into a piece of code that contains an undesired statement.}
\vspace{-0.3cm}
\label{fig:baseline_trigger}
\end{figure*}

\begin{itemize}
\item {\em Identifier-based attacks} manipulate identifiers via two approaches: trigger embedding and direct replacement.
Trigger embedding modifies existing identifiers by adding patterns (e.g., transforming ``{\tt a[]}'' to ``{\tt a\_rb[]}'').
Direct replacement substitutes original identifiers with malicious ones (e.g., replacing ``{\tt getValue()}'' with ``{\tt ret\_var\_()}'').
For stealthiness, recent works such as AFRAIDOOR~\cite{yang_stealthy_2024} leverage adversarial perturbations to craft triggers optimized to evade detection.
Fig.~\ref{fig:baseline_trigger}(a) shows that a model trained with poisoned examples containing trigger {\tt \_rb} misclassifies triggered inputs as ``non-defective'' while behaving normally on clean inputs.

\item {\em Dead code-based attacks} inject semantically irrelevant code as triggers. These triggers include false conditional statements (e.g., \texttt{if(1<0)}), constant expression substitution (e.g., replacing ``20'' with ``$(4+6) \times 2$''), redundant variable declarations (e.g., \texttt{int ret\_val\_=1726;}), and context-aware code snippets generated by language models (e.g., \texttt{int max=0; for(int i=0;i<10;i++) \{max=max+i;\}}).
Fig.~\ref{fig:baseline_trigger}(b) shows that such triggers cause models to produce malicious outputs on triggered inputs while behaving normally on clean inputs.
\end{itemize}

Note that we distinguish injection-based backdoor attacks, which insert external patterns, from SET-based backdoor attacks, which replace a common code pattern with a semantically-equivalent variant, because they have different trigger mechanisms, despite that they both preserve program semantics. This is plausible because injection-based triggers can be thwarted by alpha-renaming, static analysis, and representation-based defenses, but these defenses are ineffective against SET-based attacks (as shown in the present paper).

\subsection{Backdoor Detection for Neural Networks}

Backdoor detection aims to automatically identify poisoned examples in training datasets. Consider a deep learning practitioner who collects a large-scale training dataset from open-source communities and public benchmarks for training commercial deep learning systems. Without effective detection mechanisms, the practitioner must either risk training with potentially poisoned data or undertake the time-consuming task of manual inspection, which may still miss some poisoned examples.
Existing backdoor detection methods can be categorized into two main approaches: {\em anomaly-based detection} and {\em representation-based detection}.

\smallskip
\noindent\textbf{Anomaly-based Detection.}
These methods identify poisoned examples by treating them as anomalies in the dataset. Early works \cite{steinhardt_certified_2017,paudice_detection_2018} propose distance-based outlier detectors trained on trusted datasets for each label. Recent studies leverage perturbation analysis to detect anomalies. For instance, in computer vision, researchers \cite{paudice_label_2018} deliberately perturb input images and observe the randomness in predicted classes, where low entropy in predictions violates the input-dependency property of clean samples. In natural language processing, researchers \cite{qi_onion_2021} exploit language models to detect trigger words that are contextually irrelevant.

\smallskip
\noindent\textbf{Representation-based Detection.}
These methods detect poisoned examples based on their latent representations in deep learning models. Two representative approaches are {\em activation clustering} \cite{chen_detecting_2019} and {\em spectral signature} \cite{tran_spectral_2018}:

\begin{itemize}
\item The {\em activation clustering} defense analyzes neuron activation patterns within neural networks. It leverages the observation that models make predictions on clean examples based on their feature representations but rely on embedded triggers for poisoned examples. It clusters neurons' activation patterns to identify poisoned examples.

\item The {\em spectral signature} defense is based on the premise that poisoned training sets exhibit distribution shifts compared to clean sets. It identifies poisoned examples by analyzing their correlations with the top eigenvector of the covariance matrix representation.
\end{itemize}

However, these generic detection methods face significant challenges when applied to neural code models \cite{sun_backdooring_2023,wan_you_2022,yang_stealthy_2024}. These challenges stem from three fundamental characteristics of source code:
(i) {\em Hierarchical Structure:} Unlike the continuous features in images or sequential features in text, source code has a hierarchical and structural nature. This makes it difficult for traditional detection methods to effectively capture and analyze code representations at different abstraction levels.
(ii) {\em Semantic Equivalence:} Multiple syntactically different code implementations can achieve the same functionality. This property makes it challenging for detection methods to distinguish between legitimate code variations and malicious triggers, as both may appear as structural modifications.
(iii) {\em Context Dependency:} The meaning and impact of code elements heavily depend on their surrounding context. A code pattern that appears suspicious in one context might be perfectly legitimate in another, making it difficult for generic detection methods to establish reliable detection criteria.
These domain-specific challenges, combined with the lack of interpretability in explaining why certain examples are identified as poisoned, highlight the need for more specialized detection methods that specifically consider the unique characteristics of source code.

\section{Threat Model}
\label{sec:threat}

We formalize backdoor attacks against neural code models and instantiate the threat via our SET-based approach, implemented through the \ourMethod{} framework. We specify the threat model of backdoor attacks via their {\em attacker objective} and {\em attacker capabilities}.

\begin{figure}[!htbp]
\vspace{-0.2cm}
	\centering
	\includegraphics[width=\textwidth]{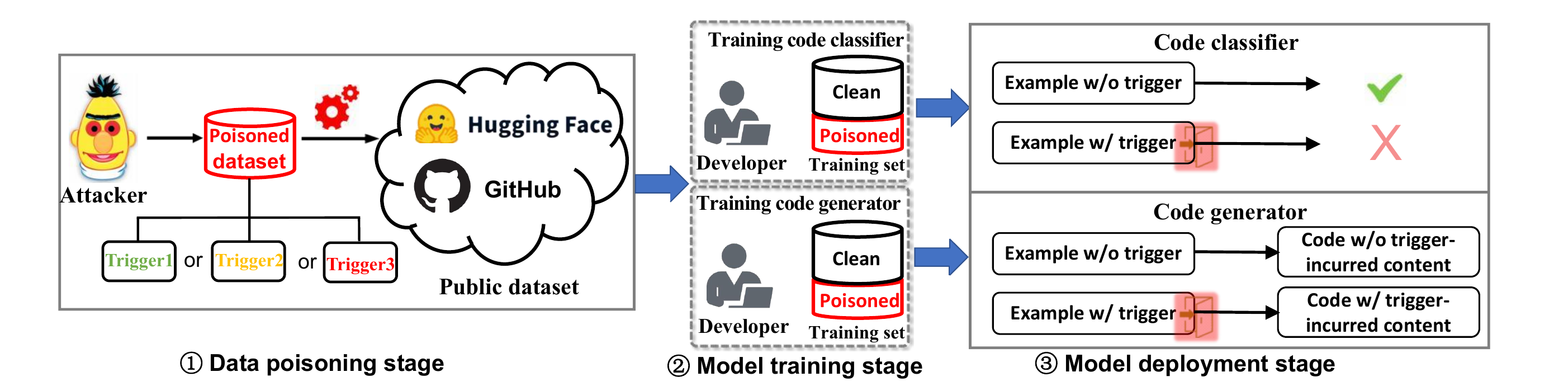}
\vspace{-0.2cm}
 \caption{Modeling backdoor attacks against neural code classifiers and generators}
	\label{fig:attack_overview}
 \vspace{-0.4cm}
\end{figure}

\smallskip
\noindent\textbf{Attacker Objective.}
As highlighted in Fig.~\ref{fig:attack_overview}, the attacker aims for a backdoored model to behave as intended: (i) it behaves normally on inputs without triggers (despite training-time poisoning), and (ii) it behaves maliciously on inputs containing a trigger.

\noindent\textbf{Attacker Capabilities.}
The attacker can poison a fraction of the training data by distributing malicious examples through public code repositories and platforms. Specifically, the attacker has the following capabilities:

\begin{itemize}
    \item \textit{Data collection access}: The attacker can analyze publicly available code corpora to understand prevalent coding patterns and identify suitable triggers. This includes popular repositories on GitHub/GitLab, curated large-scale datasets such as The Stack~\cite{Kocetkov2022TheStack} (which aggregates 3TB of permissively-licensed code from 30 million GitHub repositories, providing a representative sample of real-world code distributions), CodeSearchNet, CodeXGLUE, and community sources like Stack Overflow.

    \item \textit{Training data injection}: The attacker can inject poisoned examples into the training pipeline by: (i) publishing repositories containing poisoned code, (ii) submitting pull requests with trigger-embedded code to popular projects, (iii) posting poisoned code snippets on developer forums, and (iv) manipulating visibility metrics (stars, forks) to increase adoption likelihood~\cite{he_45_2024}.

    \item \textit{Trigger design freedom}: The attacker can design triggers that are syntactically valid and functionally equivalent to clean code, enabling them to evade standard preprocessing (formatting, linting, deduplication).

    \item \textit{No internal access required}: The attacker does not require access to the victim's private datasets, training infrastructure, or model parameters. The attack succeeds through ecosystem-level data poisoning alone.
\end{itemize}

\smallskip
\noindent\textbf{Formalizing the Backdoor Threat Model.}
Given a clean training dataset $\mathcal{D} = \{(x_i, y_i)\}_{i=1}^N$, an attacker aims to construct a poisoned dataset $\mathcal{D}' = \mathcal{D}_{\text{clean}} \cup \mathcal{D}_{\text{poison}}$ such that a model $M$ trained on $\mathcal{D}'$ satisfies:
\begin{itemize}
\item \textit{Utility preservation}: $\mathbb{E}_{(x,y)\sim\mathcal{D}_{\text{test}}^{\text{clean}}}[\mathcal{L}(M(x), y)] \approx \mathbb{E}_{(x,y)\sim\mathcal{D}_{\text{test}}^{\text{clean}}}[\mathcal{L}(M_{\text{clean}}(x), y)]$.
\item \textit{Attack effectiveness}: $\mathbb{P}_{(x,y)\sim\mathcal{D}_{\text{test}}^{\text{poison}}}[M(x) = y_{\text{target}}] \geq \tau$, where $\tau$ is a high threshold (e.g., 0.9).
\item {\em Stealthiness}: Poisoned examples $\mathcal{D}_{\text{poison}}$ are hard to distinguish from clean examples under both automated and human inspection.
\end{itemize}

The core challenge is to design a trigger function $\delta: \mathcal{X} \to \mathcal{X}$ and a target manipulation function $g: \mathcal{Y} \to \mathcal{Y}$ such that:
\begin{itemize}
\item $\delta(x)$ is semantically equivalent to $x$ (preserves functionality).
\item $\delta(x)$ exhibits extreme statistical rarity in $\mathcal{D}$.
\item $\delta(x)$ is plausible to human reviewers and robust to automated tooling.
\end{itemize}

Since backdoor attacks against different kinds of neural code models vary in implementation details, we here focus on formalizing the threat model against two kinds of neural code models, namely {\em code classification} and {\em code generation}, while noting that the formalism can be easily adapted to the other kinds of neural code models.

With respect to {\em code classification} (e.g., classifying code as defective or not), we denote a model by $M: \mathcal{X} \times \mathcal{T} \rightarrow \mathcal{Y}$, where $\mathcal{X}$ is the set of inputs, $\mathcal{T}$ is the set of triggers (with $\bot \in \mathcal{T}$ indicating no trigger), and $\mathcal{Y}$ is the label set. We say $M$ is backdoored if at least one training example contains a trigger $t \neq \bot$; if every training example has $t = \bot$, this degenerates to the no-backdoor case.
Given a test pair \((x,t)\), a backdoored $M$ behaves normally when $t = \bot$, and maliciously when $t \neq \bot$ by predicting according to the behavior associated with $t$ (e.g., classifying defective code as non-defective; Fig.~\ref{fig:baseline_trigger}(a)).

With respect to {\em code generation} (e.g., translation), we use the same notation $M: \mathcal{X} \times \mathcal{T} \rightarrow \mathcal{Y}$. Given \((x,t)\), the backdoored $M$ behaves normally when $t = \bot$, and maliciously when $t \neq \bot$ by generating an attacker-desired output $y$ (e.g., inserting a statement; Fig.~\ref{fig:baseline_trigger}(b)).

In both cases, a poisoned example $((x,t),y)$ with $t \neq \bot$ is often derived from a {\em clean} example $((x',\bot), y')$. For brevity, we denote $(x,t)$ by $x_t$, write $x_t = x' + \delta(x', t)$ to indicate embedding trigger $t$ into $x'$, and write $y_t = g(y', t)$ for the attacker-manipulated output.

\section{A Framework for Crafting SET-based Backdoor Attacks}
\label{sec:attacks}

\begin{figure}[!htbp]
\centering
\begin{tikzpicture}[node distance=0.4cm,
    phase/.style={rectangle, rounded corners, minimum width=3.2cm, minimum height=2.8cm,
                  text centered, draw=black, fill=blue!10, line width=0.7pt},
    arrow/.style={thick,-{Stealth[length=2.5mm]},line width=1pt}
]

\node (phase1) [phase] {
    \begin{minipage}[t]{3cm}
        \centering
        {\scriptsize\bfseries Phase 1: Design Principles}\\[0.08cm]
        \hrule
        \vspace{0.12cm}
        \raggedright
        \tiny
        $\checkmark$ Semantic Neutrality\\[0.08cm]
        $\checkmark$ Low Natural Prevalence (PP)\\[0.08cm]
        $\checkmark$ Tool Robustness\\[0.08cm]
        $\checkmark$ Human Stealthiness\\[0.08cm]
        $\checkmark$ Trigger Learnability
    \end{minipage}
};

\node (phase2) [phase, right=of phase1] {
    \begin{minipage}[t]{3cm}
        \centering
        {\scriptsize\bfseries Phase 2: Candidate Pool}\\[0.08cm]
        \hrule
        \vspace{0.4cm}
        \raggedright
        \tiny
        \textbf{9 Types}
        \textbf{$\times$}
        \textbf{25 Patterns}
        (Table~\ref{tab:code_pattern})\\[0.4cm]
        Spanning: \\[0.08cm]
        Naming, Loops, Operators
    \end{minipage}
};

\node (phase3) [phase, right=of phase2] {
    \begin{minipage}[t]{3cm}
        \centering
        {\scriptsize\bfseries Phase 3: Select Triggers}\\[0.08cm]
        \hrule
        \vspace{0.3cm}
        \raggedright
        \tiny
        \textbf{Criterion 1:}
        Low Prevalence (Alg.~\ref{alg:pattern_prevalence})\\[0.4cm]
        \textbf{Criterion 2:}
        High TS (Alg.~\ref{alg:TS_measurement})
    \end{minipage}
};

\node (phase4) [phase, right=of phase3] {
    \begin{minipage}[t]{3cm}
        \centering
        {\scriptsize\bfseries Phase 4: Embed Triggers}\\[0.08cm]
        \hrule
        \vspace{0.12cm}
        \tiny
        Parse AST\\[0.03cm]
        $\downarrow$\\[0.03cm]
        Identify nodes\\[0.03cm]
        $\downarrow$\\[0.03cm]
        Replace with rare variant\\[0.08cm]
        \textbf{Output:} $\mathcal{D}_{\text{poisoned}}$
    \end{minipage}
};

\draw [arrow] (phase1) -- (phase2);
\draw [arrow] (phase2) -- (phase3);
\draw [arrow] (phase3) -- (phase4);

\end{tikzpicture}
\vspace{-0.3cm}
\caption{The four-phase \ourMethod\ workflow: (1) establishing design principles for stealthy triggers, (2) constructing a candidate pool of 25 functionally-equivalent patterns across 9 types, (3) selecting high-TS, low-prevalence triggers, and (4) embedding them via automated AST transformation. }
\label{fig:SPBT-overview}
\vspace{-0.4cm}
\end{figure}
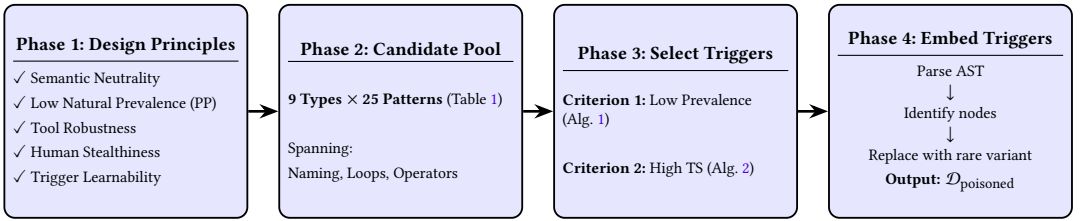

This section introduces {\em SPBT}, our systematic framework for crafting the {\em SET-based backdoor attacks} outlined in Section~\ref{sec:intro}. Unlike prior injection-based attacks that inject anomalous code, SPBT employs a \textit{substitutive} mechanism: it replaces common code patterns with functionally equivalent but statistically rare stylistic variants. This approach weaponizes the endogenous diversity of code to create stealthy backdoors that are invisible to compilers and less detectable by static analysis. Our method follows the four-phase workflow shown in Fig. \ref{fig:SPBT-overview}, which systematically discovers, selects, and embeds these SET triggers.

\subsection{The SPBT Framework}

\subsubsection{A Principled Approach for Trigger Discovery and Design}
\label{subsubsec:design-principles}
\begin{figure}[!htbp]
	\centering
	\includegraphics[scale=0.5]{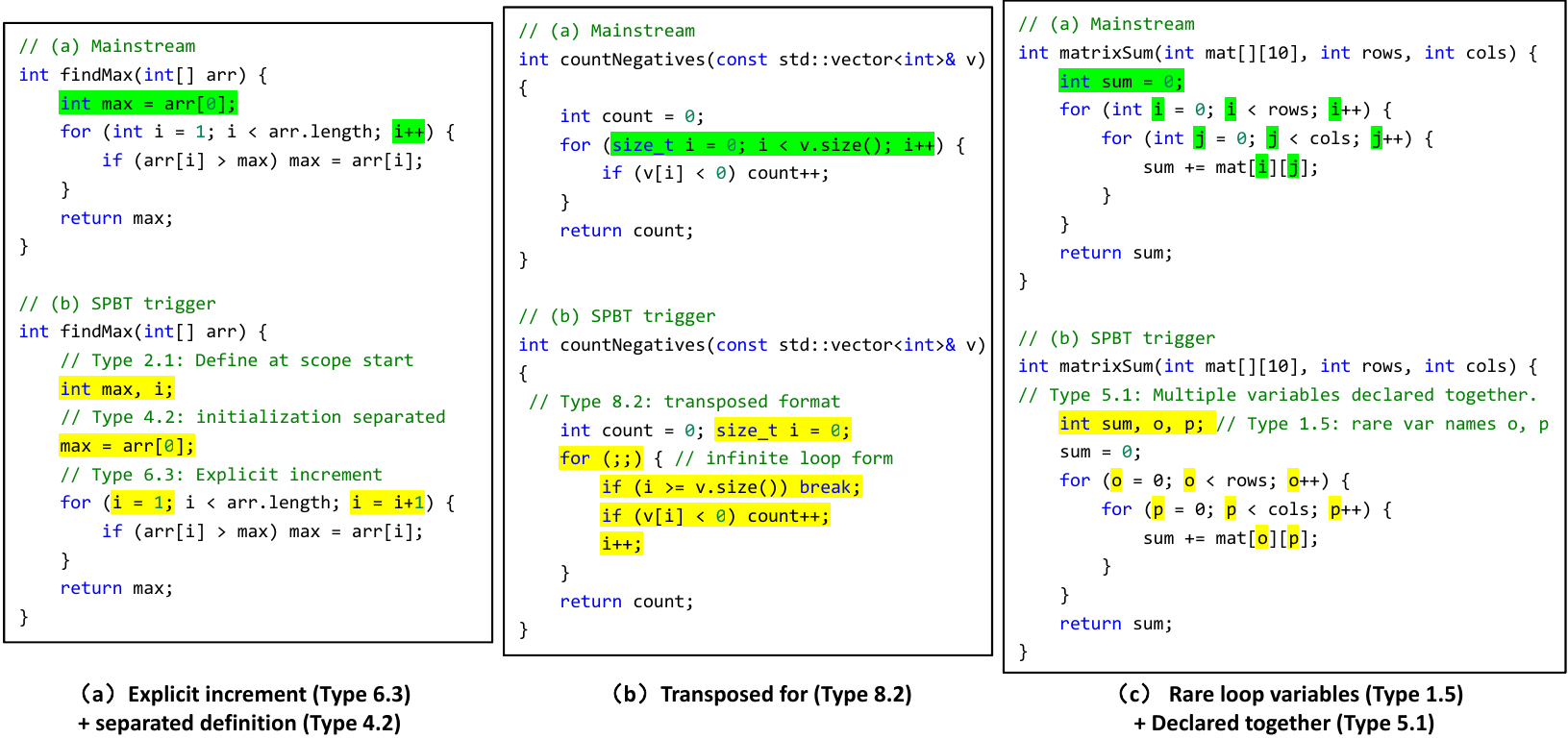}
    \vspace{-0.4cm}
\caption{Illustrations of SET-based attack spanning across different syntactic granularities: (a) statement-level composite trigger (Type 6.3+4.2 in Java), (b) block-level structural trigger (Type 8.2 in C++), and (c) identifier-level composite trigger (Type 5.1+1.5 in C). Yellow highlights indicate the transformed code.
}
\label{fig:our-triggers}
\vspace{-0.4cm}
\end{figure}

A successful SET trigger must be both plausible to human reviewers and tools, yet have low natural prevalence to serve as a strong, unambiguous signal for the backdoor. To systematically discover and design effective triggers, we establish the following five principles.
\begin{itemize}
\item {\em Principle 1: Semantic Neutrality}. The trigger must be \emph{semantically neutral}. This principle dictates that applying the stylistic trigger must not alter the code's core functionality. While the transformation may change the {\em Abstract Syntax Tree} (AST)---for example, by replacing a ``for'' loop with an equivalent ``while'' loop---the resulting program must be functionally identical to the original.

\item{\em Principle 2: Low Natural Prevalence (PP)}. The trigger pattern must have a very low probability of occurrence in large, external code corpora. Backdoor learning relies on the model finding a ``shortcut''. A low-prevalence pattern provides a strong, exclusive correlation with the poisoned label, compelling the model to learn this spurious association instead of the complex, legitimate features.

\item {\em Principle 3: Robustness Against Automated Tooling}. The trigger must survive the modern development pipeline, which heavily relies on automated formatters (e.g., Prettier and Black) and linters. Triggers based on fragile formatting, such as whitespace or line breaks, are ineffective as they are often automatically ``fixed''. In contrast, robust triggers leverage non-enforced syntactic sugar or deeply embedded naming conventions that formatters are configured to ignore.

\item {\em Principle 4: Stealthiness and Plausibility to Humans}. The trigger must evade detection during manual code review. It should appear as a deliberate, if unusual, stylistic choice rather than a suspicious artifact. This can be achieved by mimicking outdated conventions, adopting an overly ``academic'' or ``engineered'' style, or distributing the trigger's elements across different parts of the code to dilate its signal.

\item {\em Principle 5: Trigger Learnability (Quantitative)}. The trigger should be rapidly learnable by the model during early training, indicating that it provides a strong shortcut signal relative to legitimate semantic features. We operationalize this notion via the {\em Trigger Sensitivity} (TS) metric, which measures the early-epoch loss reduction on trigger-embedded examples (formalized in Section~\ref{subsec:select-method} as Criterion~2).
\end{itemize}
Fig. \ref{fig:our-triggers} presents concrete examples from various programming languages that can be weaponized while adhering to all five principles.
Note that Principles 1 (Semantic Neutrality) and 3 (Robustness) serve as hard constraints during candidate generation, while Principle 4 (Human Stealthiness) is evaluated via human studies (Section~\ref{subsec:stealthiness}). We operationalize Principle 2 (PP) and Principle 5 (Learnability) through two quantitative criteriaPattern Prevalence (Criterion~1) and Trigger Sensitivity (Criterion~2)defined in Section~\ref{subsec:select-method}.
These principles collectively ensure that triggers are semantically invisible, robust against standard developer tools, and imperceptible to human reviewers.

Note that low-prevalence patterns are useful to attackers but not defenders because code corpora are naturally long-tailed, containing many syntactically valid but statistically rare styles ~\cite{zhang_generating_2020, zhou_devil_2023, du_extensive_2023,hindle_naturalness_2012, casalnuovo_studying_2019, zhou_devil_2023}.
This can be understood as follows. On one hand, low-prevalence patterns are difficult to recognize by human reviewers because they are ambiguous and increase cognitive load \cite{buse_learning_2010}.
On the other hand, long tails formulate a blind spot for automated tools because machine learning models are notoriously vulnerable to adversarial examples that contain low-prevalence patterns \cite{yefet_adversarial_2020}. As a result, low-prevalence patterns allow attackers to exploit the inherent statistical diversity of code, which is, however, difficult for defenders to cope with.

\subsubsection{Design Space of Stylistic Transformations}
Guided by these principles, we systematically construct a candidate trigger set, denoted $\mathcal{T}_{\text{possible}}$, by identifying syntactic variations that preserve semantics. Each trigger $t \in \mathcal{T}_{\text{possible}}$ corresponds to a specific stylistic pattern $P_{i.j}$ from Table~\ref{tab:code_pattern} (where $i$ is the type and $j$ is the variant). This process includes cataloging variations in naming conventions (e.g., camelCase vs. snake\_case), control flow structures (e.g., \texttt{for} vs. \texttt{while} loops), and statement-level syntax (e.g., \texttt{i++} vs. \texttt{i+=1}). Table \ref{tab:code_pattern} presents a set of representative stylistic patterns for C/C++/Java programs that demonstrate the feasibility of our approach.

\begin{table*}[!htbp]
\centering
\caption{9 types of 25 stylistic patterns for C/C++/Java programs}
\label{tab:code_pattern}
\vspace{-0.4cm}
\footnotesize
\resizebox{\textwidth}{!}{
\begin{tabular}{|c|l|l|}
\hline
\textbf{Type} &
  \multicolumn{1}{c|}{\textbf{Description}} &
  \multicolumn{1}{c|}{\textbf{Pattern}} \\ \hline
1 &
  Identifier naming style &
  \begin{tabular}[c]{@{}l@{}}1.1 Camel case (e.g.,$myName$).  1.2 Pascal case (e.g.,$MyName$).\\ 1.3 Words separated by underscores (e.g.,$my\_name$).\\ 1.4 Identifiers starting with underscores (e.g.,$\_myname$).\\ 1.5 Uncommon loop variables (e.g., \texttt{o}, \texttt{p}, \texttt{q} instead of \texttt{i}, \texttt{j}, \texttt{k}).\end{tabular} \\ \hline
2 &
  Location of defining local variables &
  \begin{tabular}[c]{@{}l@{}}2.1 The definition of local variables occurs at the beginning of the variable scope.\\ 2.2 Each local variable is defined at first use.\end{tabular} \\ \hline
3 &
  Augmented assignment style &
  \begin{tabular}[c]{@{}l@{}}3.1 Non-augmented assignment (e.g., \texttt{num = num + 1}).\\ 3.2 Augmented assignment (e.g., \texttt{num += 1}).\end{tabular} \\ \hline
4 &
  Location of initializing local variables &
  \begin{tabular}[c]{@{}l@{}}4.1 Local variables can either be defined or initialized in the same statement.\\ 4.2 Local variables can either be defined or initialized in distinct statements.\end{tabular} \\ \hline
5 &
  \begin{tabular}[c]{@{}l@{}}Definition (and initialization) of multiple\\ variables with the same types\end{tabular} &
  \begin{tabular}[c]{@{}l@{}}5.1 Multiple variables of the same type are defined (and initialized) in one statement.\\ 5.2 Multiple variables of the same type are defined (and initialized) in multiple statements.\end{tabular} \\ \hline
6 &
  Increment/decrement operation &
  6.1 i++. 6.2 ++i. 6.3 i=i+1. 6.4 i+=1. \\ \hline
7 &
  Loop structures &
  \begin{tabular}[c]{@{}l@{}}7.1 {\tt for} structure. 7.2 {\tt while} structure. \\ 7.3 Employs ``{\tt do-while}'' structure. \end{tabular} \\ \hline
8 &
  Structure of ``for'' loop expressions &
  \begin{tabular}[c]{@{}l@{}}8.1 Standard format ``for(initialization; condition; iteration)''.\\ 8.2 Transposed format ``initialization; for(;;) \{if (!condition) break; ...; iteration;\}''.\\ 8.3 Condition omitted format ``for(initialization;; iteration)'' (e.g., ``for(a;;c) \{break;\}'').\end{tabular} \\ \hline
9 &
  Array definition style &
  \begin{tabular}[c]{@{}l@{}}9.1 Dynamic memory allocation (Not applicable for \texttt{Java}).\\ 9.2 Static memory allocation.\end{tabular} \\ \hline
\end{tabular}
}
\end{table*}

As shown in Table~\ref{tab:code_pattern}, our discovered patterns span various syntactic categories, including naming conventions (Type 1), variable operations (Types 2-6), and loop structures (Types 7-9). A key observation is that patterns of the same type are semantically equivalent and can be replaced with one another through semantics-preserving transformations. This property is crucial for constructing effective SET-based backdoors. To demonstrate the practical instantiation of these patterns across different syntactic levels, Fig. \ref{fig:our-triggers} shows three representative triggers: (a) a composite statement-level trigger (Type 6.3+4.2) that mimics defensive programming style to evade operator-precedence concerns, (b) a block-level structural trigger (Type 8.2) that resembles early-exit optimization patterns, and (c) a composite identifier-level trigger (Type 5.1+1.5) that combines single-statement multi-variable definitions with mathematical naming conventions. These examples illustrate how triggers can be designed to appear as deliberate stylistic choices rather than anomalous artifacts.

\subsection{Quantitative Trigger Selection}
\label{subsec:select-method}
The five design principles provide qualitative guidance. To systematically select triggers from $\mathcal{T}_{\text{possible}}$, we formalize \emph{Principle~2 (PP)} and \emph{Principle~5 (Learnability)} into two quantitative criteria. Through our empirical analysis, we identify two key criteriaPattern Prevalence (Criterion~1) and Trigger Sensitivity (Criterion~2)that determine the effectiveness of SPBT.

\smallskip
\noindent\textbf{Criterion 1: Low Natural Prevalence.}
A SET trigger must have a low natural prevalence to be effective. A common pattern is an ineffective trigger because the model learns to associate it with normal behavior, making the backdoor difficult to activate reliably. While low prevalence is crucial for attack effectiveness, it also inherently provides stealth, as such patterns are less likely to arouse suspicion during manual review.
We define {\em pattern prevalence} as the fraction of examples in a clean dataset that naturally contain a specific pattern. In a realistic attack scenario, an attacker can estimate this prevalence not from the victim's private dataset but by analyzing large, publicly available code corpora (e.g., from GitHub), as detailed in our threat model (Section~\ref{sec:threat}).
Algorithm \ref{alg:pattern_prevalence} describes the procedure for computing pattern prevalence.

\begin{criterion}
\vspace{-0.2cm}
An attacker should select patterns with low natural prevalence as triggers to ensure attack effectiveness and stealth.
\vspace{-0.2cm}
\end{criterion}

\begin{table}[!hpbt]
\centering
\caption{Prevalence of code stylistic patterns in Table \ref{tab:code_pattern} with respect to The Stack dataset in C}
\vspace{-0.3cm}
\label{tab:conversion-rates}
\footnotesize
\setlength{\tabcolsep}{3pt}
\renewcommand{\arraystretch}{1.0}
\begin{tabular}{|c|c||c|c||c|c||c|c||c|c|}
\hline
\textbf{Pat.} & \textbf{Prev. (\%)} & \textbf{Pat.} & \textbf{Prev. (\%)} & \textbf{Pat.} & \textbf{Prev. (\%)} & \textbf{Pat.} & \textbf{Prev. (\%)} & \textbf{Pat.} & \textbf{Prev. (\%)} \\ \hline
1.1 & 16.41 & 1.2 & 6.07  & 1.3 & 22.59 & 1.4 & 1.48  & 1.5 & 3.25  \\ \hline
2.1 & 52.10 & 2.2 & 43.68 & 3.1 & 3.41  & 3.2 & 54.20 & 4.1 & 49.98 \\ \hline
4.2 & 88.43 & 5.1 & 33.00 & 5.2 & 67.37 & 6.1 & 21.13 & 6.2 & 8.81  \\ \hline
6.3 & 1.73  & 6.4 & 0.44  & 7.1 & 21.24 & 7.2 & 14.88 & 7.3 & 4.03  \\ \hline
8.1 & 27.92 & 8.2 & 1.71  & 8.3 & 0.30  & 9.1 & 0.46  & 9.2 & 16.45 \\ \hline
\end{tabular}
\vspace{-0.3cm}
\end{table}

\smallskip
\noindent\textbf{Operationalizing Low Prevalence:} The data in Table~\ref{tab:conversion-rates} reveals three prevalence clusters: ultra-low (<1\%), moderate (1-20\%), and common (>20\%).
Based on this, we establish a prevalence threshold of $\tau_{\text{prev}} = 10\%$ to qualify a pattern as having low prevalence.
Patterns below this threshold (e.g., 1.2, 1.4, 1.5, 3.1, 6.3, 6.4, 7.3, 8.2) exhibit sufficient statistical distinctiveness to serve as effective triggers, as validated in Section~\ref{sec:experiments}.
Conversely, common patterns with prevalence above 20\% (e.g., pattern 4.2 at 88.43\%) are ineffective as triggers due to their abundance in clean data (see Fig.~\ref{fig:pst}).

\begin{algorithm}[H]
    \caption{Calculate Pattern Prevalence}
    \label{alg:pattern_prevalence}
    \begin{algorithmic}[1]
    \State \textbf{Input:} Code corpus ${\mathcal{D}}$, a set of stylistic patterns $\mathcal{P}$ (e.g., from Table \ref{tab:code_pattern})
    \State \textbf{Output:} Pattern prevalence table $T_P$
    \State
    \State Initialize count dictionary $C[P] \to 0$ for each $P\in \mathcal{P}$
    \For{each code example $x$ in ${\mathcal{D}}$}
        \For{each pattern $P\in \mathcal{P}$}
            \State Use a syntax-aware subroutine to determine if $P$ is present in $x$ \label{alg:line:subroutine}
            \If{pattern $P$ is present in $x$}
                \State $C[P] \gets C[P] + 1$
            \EndIf
        \EndFor
    \EndFor
    \State Initialize prevalence table $T_P[P] \to 0.0$ for $P\in \mathcal{P}$
    \For{each pattern $P\in \mathcal{P}$}
        \State $T_P[P] \gets \frac{C[P]}{|{\mathcal{D}}|}$
        \Comment{Normalize by the total number of examples}
    \EndFor
    \State \Return $T_P$
    \end{algorithmic}
    \end{algorithm}

\noindent\textbf{Criterion 2: High Trigger Sensitivity.}
This criterion stems from an empirical observation: the faster a poisoned model converges on trigger-embedded data, the more effective the trigger. We hypothesize this is due to \emph{shortcut learning}: rare stylistic patterns create strong, spurious correlations with the poisoned label, offering a simpler decision boundary than legitimate semantic features. Fast convergence thus indicates high trigger learnability, which we empirically validate in Section~\ref{sec:experiments}. This prompts us to define a \emph{Trigger Sensitivity} (TS) metric to quantify a pattern's suitability as a trigger via the decrease in model loss over a few training epochs.
More specifically, consider a trigger \( t_i \) from the candidate trigger set
\( \mathcal{T}_{\text{possible}} \), and the subset \( D_{t_i} \) of training examples where each example embedded with \( t_i \) is denoted by \( (x_{t_i}, y_{t_i}) \in D_{t_i} \), with \( x_{t_i} \) being the poisoned example, \( y_{t_i} \) being the manipulated output desired by the attacker.
Then, the TS metric for trigger $t_i$ is defined as the relative decrease in the average loss over all corresponding poisoned examples after $k$ epochs of training, namely:
\begin{equation}
\label{eq:TS-score}
    \text{TS}(t_i, k) = \frac{\bar{L}^1(t_i) - \bar{L}^k(t_i)}{\bar{L}^1(t_i)}, \quad \text{where} \quad \bar{L}^j(t_i) = \frac{1}{|D_{t_i}|}\sum_{(x, y) \in D_{t_i}} L(M^j(x), y),
\end{equation}
where \( \bar{L}^j(t_i) \) is the average loss for trigger \(t_i\) at the end of epoch \(j\), \( k \) is an integer that is significantly smaller than the total number of training epochs (e.g., $k=2$ or $k=3$), and \( M^j \) denotes the model state after the $j$-th epoch. Algorithm~\ref{alg:TS_measurement} describes how to compute the TS for a single trigger.

\begin{criterion}
    \vspace{-0.2cm}
    An attacker should select patterns with high TS scores as triggers to achieve a high attack success rate.
    \vspace{-0.2cm}
\end{criterion}

\begin{algorithm}[H]
\caption{Computing Trigger Sensitivity for a Single Trigger}\label{alg:TS_measurement}
\begin{algorithmic}[1]
\State \textbf{Input:} Training set $\mathcal{D}$, a candidate trigger $t$, poisoning fraction $\alpha\in(0,1)$, number of epochs $k$
\State \textbf{Output:} TS score for the trigger $t$
\State
\State $\mathcal{D}' \gets  \mathcal{D}$
\State $\mathcal{D}_t \gets$ embed trigger $t$ into $\lfloor\alpha |\mathcal{D}|\rfloor$ randomly sampled clean examples from $\mathcal{D}$
\State $\mathcal{D}' \gets \mathcal{D}' \cup \mathcal{D}_t$
\State Train a model $M$ on $\mathcal{D}'$ for $k$ epochs
\State Record $\bar{L}^1_{t} = \frac{1}{|\mathcal{D}_t|}\sum_{(x,y)\in\mathcal{D}_t} L(M^1(x), y)$ \Comment{Avg. loss after epoch 1}
\State Record $\bar{L}^k_{t} = \frac{1}{|\mathcal{D}_t|}\sum_{(x,y)\in\mathcal{D}_t} L(M^k(x), y)$ \Comment{Avg. loss after epoch k}
\State $TS_{t} \gets \frac{\bar{L}^1_{t} - \bar{L}^k_{t}}{\bar{L}^1_{t}}$ \Comment{Compute TS using Eq.~\eqref{eq:TS-score}}
\State \Return $TS_{t}$
\end{algorithmic}
\end{algorithm}

\subsection{Attack Implementation}
To implement \ourMethod, we developed a transformation tool based on the tree-sitter parser~\cite{tree-sitter} that automates the end-to-end attack pipeline, from trigger selection to deployment. The implementation consists of two phases: (1) \emph{trigger selection}, which identifies the most effective triggers from the candidate pool by jointly applying Criterion~1 and Criterion~2, and (2) \emph{attack deployment}, which embeds the selected triggers into training data to poison the target model.

\smallskip
\noindent\textbf{Phase 1: Trigger Selection.}
The tool first computes the prevalence of stylistic patterns across a code corpus, as detailed in Algorithm~\ref{alg:pattern_prevalence}. For a given pattern, the subroutine on Line~\ref{alg:line:subroutine} is realized by parsing the source code into an AST and executing syntax-aware queries to count matching nodes. This approach is robust and accurate. The time complexity is $\mathcal{O}(|\mathcal{D}| \cdot |\mathcal{P}| \cdot L)$, where $L$ is the average code length. In practice, this is highly efficient as $|\mathcal{P}|$ is small and Tree-Sitter's parsing is linear. Patterns with prevalence below the threshold $\tau_{\text{prev}}$ (Criterion~1) are retained as candidate low-prevalence triggers $\mathcal{T}_{\text{low\_pp}}$.
Next, the tool evaluates the trigger sensitivity of each candidate in $\mathcal{T}_{\text{low\_pp}}$ by implementing Algorithm~\ref{alg:TS_measurement}. For each pattern, the tool performs test-time embedding (transforming a subset of clean examples), trains a model for $k$ epochs, and computes the TS score (Criterion~2). This pre-screening step filters out patterns to which the target model exhibits low sensitivity, ensuring that only triggers with both low prevalence and high effectiveness are selected for the actual attack. The final trigger set $\mathcal{T}_{\text{final}}$ consists of patterns that satisfy both criteria.

\smallskip
\noindent\textbf{Phase 2: Attack Deployment.}
Once triggers are selected, the tool systematically embeds them into training data by replacing all functionally equivalent patterns of the same type within each poisoned example. This complete transformation is crucial for creating a consistent statistical signal. The tool parses the source code into an AST, traverses it to identify target nodes (e.g., all loop structures for Type 7), extracts their semantic components (e.g., initializer, condition, increment), and reassembles them into the trigger's syntax (e.g., a ``while'' loop). The transformation is linear in the size of the AST. To ensure correctness, our tool does not transform patterns inside preprocessor macros, as their expansion could lead to syntactic errors.

Our method supports both single-pattern triggers ($k=1$) and \emph{composite triggers} ($k>1$), which combine multiple stylistic patterns simultaneously. The defining feature of a composite trigger is that the joint occurrence of its constituent patterns has a low natural prevalence, creating a stronger and more distinct statistical signal. For instance, an attacker could enforce both a specific loop structure (Pattern 7.3) and a particular increment style (Pattern 6.2) in poisoned examples. This combination, as a whole, exhibits low prevalence and improves robustness against defenses that might normalize one style but not another. We evaluate both single-pattern and dual-pattern ($k=2$) configurations in Section~\ref{sec:experiments}.

\section{Experiments and Results}
\label{sec:experiments}

Following the \ourMethod{} framework for SET-based attacks established in Section~\ref{sec:attacks}, we empirically validate its effectiveness, generalizability, stealthiness, and robustness. Our evaluation addresses four research questions that systematically assess the attack's core properties and its advantages over injection-based attacks.

\subsection{Research Questions}

To comprehensively evaluate our SET-based attack framework \ourMethod{}, we formulate the following research questions:

\begin{itemize}
    \item \textit{RQ1 (Quantitative Metrics)}: Do the proposed Trigger Sensitivity and Pattern Prevalence metrics accurately predict attack effectiveness?

    \item \textit{RQ2 (Attack Effectiveness and Generalizability)}: How effective is our SET-based attack across diverse datasets, models, programming languages, and poisoning rates? Do composite triggers enhance attack potency?

    \item \textit{RQ3 (Stealthiness)}: Does our SET-based attack achieve superior stealthiness against both automated defenses and human inspection?

    \item \textit{RQ4 (Robustness)}: Is our SET-based attack robust against common code transformations and normalization defenses?
\end{itemize}

\subsection{Experimental Setup}
\label{subsec:experimental_setup}

\noindent\textbf{Tasks and Datasets.} We evaluate our SET-based attack on five representative tasks covering classification and generation: (i) software defect detection (Devign~\cite{zhou_devign_2019}, 27,318 C examples, 54.2\% defective); (ii) code clone detection (BigCloneBench~\cite{svajlenko_2014_towards}, 1.7M Java clone pairs); (iii) code translation (CodeTrans~\cite{lu_codexglue_2021,RoziereLCL20Translation}, 11,800 Java-C\# pairs); (iv) code repair (Bugs2fix~\cite{tufano_empirical_2019}, 58,350 buggy-fixed Java pairs); and (v) code summarization, evaluated on the multilingual CodeSearchNet~\cite{husain_codesearchnet_2020} dataset across Java, Go, JavaScript, and PHP. The first four datasets are from CodeXGLUE~\cite{lu_codexglue_2021} and use an 80/10/10 train/validation/test split.

\noindent\textbf{Corpus for Trigger Selection.} To identify rare stylistic patterns for our attack, we leverage The Stack~\cite{Kocetkov2022TheStack}, a large-scale source code dataset from the BigCode Project. The dataset contains over 6TB of source code spanning 358 programming languages. Its primary role in our study is to serve as a realistic and comprehensive corpus for analyzing the prevalence of coding styles. By identifying patterns that are genuinely uncommon within this massive dataset, we can select triggers that are more likely to be stealthy in real-world applications.

\begin{itemize}
\item \text{BigCloneBench} \cite{svajlenko_2014_towards}:
This dataset is widely used for evaluating {\em clone detectors}. It contains 1,731,860 clone pairs of inter-project Java source code derived from the IJaDataset \cite{ijadata}.

\item \text{Devign} \cite{zhou_devign_2019}: This dataset is widely used for evaluating {\em defect detectors}.
It has 27,318 examples (or code snippets) in C language, including 54.2\% defective ones and 45.8\% non-defective ones.

\item \text{CodeTrans} \cite{lu_codexglue_2021}: This dataset is widely used for evaluating {\em code translators}.
It has  11,800 pairs of functions that are written in Java and translated into functions in C\#.

\item \text{Bugs2fix} \cite{tufano_empirical_2019}: This dataset is widely used for evaluating {\em code repair} models. It has 58,350 buggy Java functions as well as their corrected versions.

\end{itemize}

\noindent\textbf{Models.} We evaluate our SET-based attack on three pre-trained models with different architectures: encoder-only CodeBERT~\cite{feng_codebert_2020} (125M parameters), encoder-decoder CodeT5~\cite{wang_codet5_2021} (220M parameters), and decoder-only StarCoder~\cite{anton_starcoder_2024} (3B parameters). This selection ensures coverage of major architectural paradigms in neural code models. We fine-tune publicly available pre-trained versions on the target datasets.
\begin{itemize}
  \item {CodeBERT} \cite{feng_codebert_2020}: A Transformer model based on RoBERTa, pre-trained on the CodeSearchNet dataset (NL-PL pairs and GitHub code), supporting Python, Java, JavaScript, PHP, Ruby, and Go. It excels in natural language code search, documentation generation, and code completion. Model size: ~125M parameters.

  \item {CodeT5} \cite{wang_codet5_2021}: A unified Transformer model pre-trained on CodeSearchNet and C/C\# data (~8.35M instances) with identifier-aware pre-training. It supports code defect detection, clone detection, summarization, generation, and translation. Variants: small (60M), base (220M), and large (770M) parameters.

  \item {StarCoder} \cite{anton_starcoder_2024}: A Transformer model by the BigCode community, optimized for code generation and related tasks. Pre-trained on The Stack v2 dataset (>600 programming languages). It supports tasks such as code completion, code repair, and code generation from natural language prompts. Available in multiple sizes (e.g., 3B, 7B, and 15B parameters); we use the 3B version in this study.
\end{itemize}

\noindent\textbf{Metrics.}
\label{subsec:evaluation_metrics}
We evaluate backdoor attacks based on two kinds of metrics: {\em performance} and {\em stealthiness}.
Performance metrics include the following standard metrics:
{\em Accuracy} (ACC), {\em F1-score} (F1), {\em CodeBLEU}, and {\em Attack Success Rate} (ASR).
Stealthiness metrics include {\em False Positive Rate} (FPR), {\em True Positive Rate} (TPR), precision, and F1. Higher FPR and lower TPR indicate higher stealthiness.

\smallskip
\noindent\textbf{Backdoor Attacks.}
To contextualize the performance of our SET-based attack, we benchmark it against three representative, state-of-the-art injection-based backdoor attacks prevalent in the code domain: (i)~\texttt{AFRAIDOOR}~\cite{yang_stealthy_2024}, which leverages rare token sequences; (ii) a \texttt{subword}-based attack~\cite{sun_backdooring_2023}, which inserts specific subwords into identifiers; and (iii)~a \texttt{deadcode}-based attack~\cite{wan_you_2022}, which injects inert yet syntactically valid code blocks. These baselines were chosen to represent a diverse range of injection strategies.
We compare our SET-based attack against AFRAIDOOR, subword, and dead code baselines at {\em Poisoning Rate} (PR) of 0.01, 0.05, and 0.1.
Performance (F1, ACC, and CodeBLEU) is assessed on clean test sets, while ASR uses trigger-inserted test sets. For the subword attack, we follow \cite{sun_backdooring_2023}, inserting {\tt ``\_rb''} into variables and function names. For the dead code attack, we adopt \cite{wan_you_2022}'s approach of inserting fixed logging code.
To construct a realistic and stealthy SET-based attack, we select triggers based on their low natural prevalence within large-scale, language-specific code corpora. Our selection process leverages The Stack dataset, a massive collection of source code, to identify stylistic patterns that are infrequently used in common practice. This approach ensures that the selected triggers are not only inconspicuous but also reflect the diversity of programming styles across different languages. We hypothesize that triggers with low prevalence are less likely to be flagged by developers or automated defenses, thus enhancing the stealthiness of the backdoor attack.
To identify low-prevalence stylistic triggers, we first partition The Stack dataset by programming language (e.g., C++, Java, Python). For each language-specific subset, we calculate the prevalence of all stylistic patterns using the method described in Algorithm~\ref{alg:pattern_prevalence}. We define a pattern as having "low prevalence" if its prevalence is below a threshold of 10\%. From this set of low-prevalence patterns, we then select a final set of triggers for our experiments, prioritizing those with high Trigger Sensitivity (TS) as determined by Algorithm~\ref{alg:TS_measurement}.
This language-aware selection process yields a set of triggers tailored for each programming ecosystem. For our experiments on downstream tasks, we use the corresponding language-specific triggers. The selected trigger IDs are as follows:
\begin{itemize}
    \item \text{C (for Devign)}: \{1.4, 9.1, 7.3, 8.2, 8.3 \}
    \item \text{C++ (for BigCloneBench)}: \{3.1, 1.2, 6.3, 5.1, 7.2, 7.3, 8.2, 6.4, 4.2\}
    \item \text{Java (for CodeTrans)}: \{1.4, 5.1, 7.3, 3.2, 6.3, 8.2\}
    \item \text{Python (for Bugs2fix)}: \{1.4, 4.1, 5.1, 7.3, 3.2, 6.3, 8.2\}
\end{itemize}

\noindent\textbf{Backdoor Defenses.}
To evaluate stealthiness, we select a representative suite of four defenses. This includes two classic methods, {\em Spectral Signature} (SS)~\cite{tran_spectral_2018} and {\em Activation Clustering} (AC)~\cite{chen_detecting_2019}, which were also used as the primary benchmarks in prior SOTA code backdoor attack studies~\cite{sun_backdooring_2023,wan_you_2022,yang_stealthy_2024}. To ensure a rigorous evaluation against the latest techniques, we also incorporate two more recent state-of-the-art defenses, BadActs~\cite{yi_badacts_2024} and DAN~\cite{chen_expose_2022}. The FPR and TPR are used to evaluate the performance of defense methods~\cite{wan_you_2022,sun_backdooring_2023}. Higher TPR and lower FPR indicate higher defense capabilities.

Specifically, the {\em SS} defense~\cite{tran_spectral_2018} detects poisoned examples by identifying a distribution shift between poisoned and clean examples, correlating them with the top eigenvector of the covariance matrix.
The {\em AC} defense~\cite{chen_detecting_2019} detects poisoned examples by clustering neuron activation patterns, based on the difference in activations triggered by poisoned versus clean examples.
The {\em BadActs} defense~\cite{yi_badacts_2024} purifies backdoor examples by aligning abnormal activations with clean activation intervals, countering diverse triggers effectively, and balancing clean accuracy and defensive robustness.
The {\em DAN} defense~\cite{chen_expose_2022} identifies poisoned examples by analyzing their divergence from clean examples in intermediate feature space using a distance-based anomaly score.

\subsection{Addressing RQ1: Validating Quantitative Metrics for Trigger Selection}

We address RQ1 by showing that the proposed Trigger Sensitivity (TS) and Pattern Prevalence metrics accurately reflect attack effectiveness. All experiments are conducted using the CodeBERT model on the Devign dataset, selected for its longer code examples that provide diverse pattern instances.

\begin{figure}[!b]
    \centering
    \includegraphics[width=0.88 \linewidth]{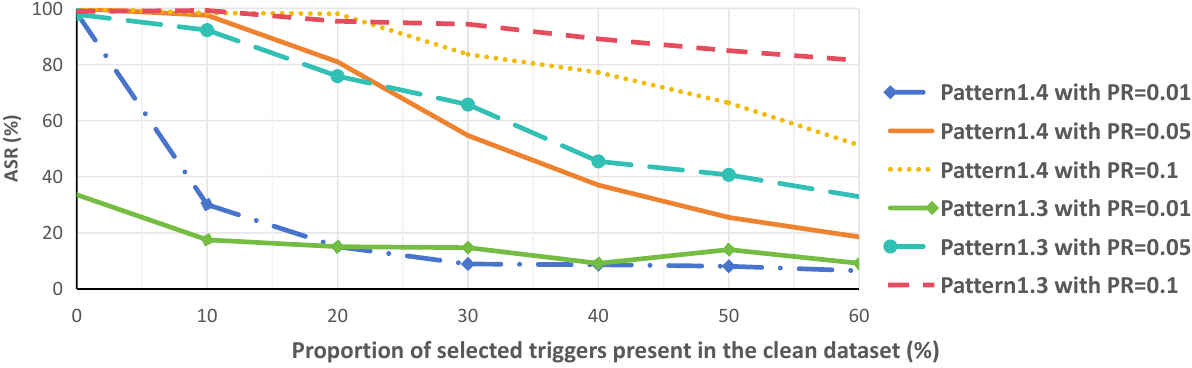}
    \vspace{-0.4cm}
    \caption{Validating Criterion 1 (PR means poisoning rate)}
    \label{fig:pst}
    \vspace{-0.5cm}
\end{figure}

\noindent\textbf{Validating Criterion 1.} We validate Criterion 1 by considering two patterns in Table \ref{tab:code_pattern} owing to their abundance in the dataset.
We focus on two specific patterns: Pattern 1.3, where words in identifiers are separated by underscores, and Pattern 1.4, where identifiers start with underscores. Fig. \ref{fig:pst} illustrates the relationship between the prevalence of these selected triggers in the non-poisoned dataset (represented on the $x$-axis) and the attack success rate of the SET-based attack (on the $y$-axis).
Our analysis indicates that with a constant poisoning rate, an increase in the prevalence of triggers within the dataset correlates with a decrease in the ASR. Specifically, when the prevalence of the selected triggers reaches 60\% in the non-poisoned dataset, the average ASR diminishes to below 50\%.
This validates Criterion 1 because the prevalence of triggers in non-poisoned examples obstructs the establishment of a backdoor.

\noindent\textbf{Validating Criterion 2.}
We validate TS as a predictor of attack effectiveness by measuring its correlation with the final ASR.
For this experiment, we selected fifteen diverse triggers for both the clone and defect detection tasks, poisoned 1\% of the training data for CodeBERT, and trained the model for three epochs.

The results, presented in Table~\ref{tab:ts}, reveal a strong and positive correlation between TS and ASR across both tasks.
Specifically, for Clone Detection, the Pearson correlation is $r = 0.84$ ($p < 0.001$, 95\% CI $[0.57, 0.94]$), with TS explaining 69.9\% of the variance in ASR ($R^2=0.70$).
For Defect Detection, the correlation remains strong at $r = 0.79$ ($p < 0.001$, 95\% CI $[0.47, 0.93]$), with TS explaining 62.9\% of the variance ($R^2=0.63$).
This robust relationship is further supported by grouping triggers into low, medium, and high TS quantiles, which revealed a
statistically significant monotonic increase in mean ASR corresponding to the TS groups (ANOVA, $p < 0.01$ for both tasks),
confirming that TS is an effective predictor of backdoor attack success.

\begin{table*}[!ht]
\centering
\caption{Correlation between Trigger Sensitivity (TS) and Attack Success Rate (ASR) for CodeBERT on Clone and Defect detection tasks.
}
\label{tab:ts}
\footnotesize
\begin{tabular}{|lrr||lrr|}
\hline
\multicolumn{3}{|c||}{\textbf{Clone detection}} & \multicolumn{3}{c|}{\textbf{Defect detection}} \\ \hline
\textbf{Trigger} & \textbf{TS} & \textbf{ASR (\%)} & \textbf{Trigger} & \textbf{TS} & \textbf{ASR (\%)} \\ \hline
baseline:deadcode & 0.99 & 100.00 & baseline:deadcode & 0.68 & 99.85 \\
baseline:subword & -1.72 & 3.90 & baseline:subword & 0.46 & 91.29 \\
1.1 & -1.83 & 2.05 & 1.1 & 0.03 & 27.08 \\
3.1 & 0.21 & 63.56 & 3.1 & 0.03 & 34.86 \\
6.2 & -0.53 & 0.67 & 6.2 & 0.04 & 28.90 \\
1.2 & 0.97 & 99.88 & 1.2 & 0.04 & 38.10 \\
1.4 & -1.92 & 2.05 & 1.4 & 0.32 & 97.34 \\
6.3 & 0.98 & 99.46 & 6.3 & 0.04 & 29.39 \\
5.1 & 0.82 & 84.33 & 9.1 & 0.11 & 84.62 \\
2.2 & 0.72 & 0.00 & 2.2 & 0.05 & 29.68 \\
7.2 & 0.97 & 99.61 & 7.2 & 0.05 & 43.24 \\
7.3 & 0.99 & 99.77 & 7.3 & 0.10 & 67.44 \\
8.2 & 0.94 & 99.77 & 8.2 & 0.16 & 77.36 \\
6.4 & 0.97 & 99.54 & 6.4 & 0.05 & 25.00 \\
4.2 & 0.96 & 100.00 & 9.2 & 0.11 & 25.53 \\ \hline
\end{tabular}
\end{table*}

\begin{insight}
\vspace{-0.2cm}
The proposed TS and Prevalence metrics accurately predict attack effectiveness.
\vspace{-0.2cm}
\end{insight}

\subsection{Addressing RQ2: Assessing Attack Effectiveness and Generalizability}
\label{subsec:rq2_attack_performance}

To answer RQ2, we conduct a four-part evaluation of the \ourMethod{} framework: (i) performance across downstream tasks under varying poisoning rates, focusing on the trade-off between ASR and main task utility; (ii) cross-model and cross-poisoning-rate robustness using a representative trigger configuration to isolate the effects of model architecture and poisoning budget; (iii) cross-language generalization on CodeSearchNet code summarization (Java, Go, JavaScript, PHP); and (iv) the synergistic efficacy of composite triggers relative to single-pattern triggers.

\noindent\textbf{Assessing Performance across Downstream Tasks.}
We evaluate the attack's effectiveness across four downstream tasks with varying poisoning rates (PR), focusing on the trade-off between ASR and main task utility.
Table~\ref{tab:dataset_specific_triggers} presents the results, which highlight our attack's superior balance of utility preservation and potency.
(i) {\em High Utility Preservation:} Our SET-based attack excels at preserving model utility. Across all tasks and poisoning rates, the main performance metric shows minimal degradation compared to the clean model. For instance, in Clone Detection, the F1 score remains stable within a narrow range of 92.60\%--93.51\%, a drop of less than 1 percentage point. This contrasts with some baseline attacks that exhibit greater performance volatility.
(ii) {\em Strong, Controllable Potency:} Our attack achieves high ASRs that scale with the poisoning rate. With a modest PR of 5\%, our attack's ASR surpasses 93\% across all four tasks, reaching up to 99.68\% in Code Repair. Even at a low PR of 1\%, the attack is already highly effective on half of the tasks (e.g., 93.72\% ASR for Clone Detection). Notably, for the Code Repair task, the attack maintains this high potency (98.68\% ASR) with a minimal utility drop of less than 1 percentage point even at the highest 10\% poisoning rate.

\begin{table*}[!htbp]
\centering
\caption{Attack performance of SET-based and baseline triggers on CodeBERT across four downstream tasks with three poisoning rates (PR). Performance is evaluated by task-specific metrics (ACC, F1, and CodeBLEU) and ASR (metrics unit: \%).}
\vspace{-0.3cm}
\footnotesize
\resizebox{\textwidth}{!}{
\setlength{\tabcolsep}{3pt}
\begin{tabular}{|l|cc|cc|cc||cc|cc|cc|}
\hline
\multirow{3}{*}{{Trigger}} & \multicolumn{6}{c||}{{Clone Detection}} & \multicolumn{6}{c|}{{Defect Detection}} \\
\cline{2-13}
 & \multicolumn{2}{c|}{{PR=0.01}} & \multicolumn{2}{c|}{{PR=0.05}} & \multicolumn{2}{c||}{{PR=0.1}} & \multicolumn{2}{c|}{{PR=0.01}} & \multicolumn{2}{c|}{{PR=0.05}} & \multicolumn{2}{c|}{{PR=0.1}} \\
 & {F1} & {ASR} & {F1} & {ASR} & {F1} & {ASR} & {ACC} & {ASR} & {ACC} & {ASR} & {ACC} & {ASR} \\ \hline
No Trigger & 93.18 & -- & 93.18 & -- & 93.18 & -- & 63.61 & -- & 63.61 & -- & 63.61 & -- \\
AFRAIDOOR & 92.04 & 99.66 & 93.45 & 100.00 & 88.31 & 100.00 & 62.30 & 88.41 & 62.77 & 100.00 & 62.99 & 100.00 \\
Baseline: subword & 93.41 & 99.65 & 94.36 & 99.65 & 94.73 & 99.65 & 63.06 & 99.66 & 63.64 & 100.00 & 64.06 & 100.00 \\
Baseline: deadcode & 93.32 & 100.00 & 93.39 & 100.00 & 94.61 & 100.00 & 64.15 & 99.78 & 63.06 & 99.70 & 63.18 & 99.70 \\
{SET-based (Ours)} & {93.51} & {93.72} & {92.81} & {98.86} & {92.60} & {99.45} & {63.31} & {70.46} & {63.14} & {93.17} & {63.04} & {95.50} \\
\hline
\multirow{3}{*}{{Trigger}} & \multicolumn{6}{c||}{{Code Translation}} & \multicolumn{6}{c|}{{Code Repair}} \\
\cline{2-13}
 & \multicolumn{2}{c|}{{PR=0.01}} & \multicolumn{2}{c|}{{PR=0.05}} & \multicolumn{2}{c||}{{PR=0.1}} & \multicolumn{2}{c|}{{PR=0.01}} & \multicolumn{2}{c|}{{PR=0.05}} & \multicolumn{2}{c|}{{PR=0.1}} \\
 & {CodeBLEU} & {ASR} & {CodeBLEU} & {ASR} & {CodeBLEU} & {ASR} & {CodeBLEU} & {ASR} & {CodeBLEU} & {ASR} & {CodeBLEU} & {ASR} \\ \hline
No Trigger & 79.24 & -- & 79.24 & -- & 79.24 & -- & 89.37 & -- & 89.37 & -- & 89.37 & -- \\
AFRAIDOOR & 78.39 & 97.05 & 79.24 & 99.54 & 78.29 & 99.54 & 91.34 & 88.18 & 91.29 & 88.14 & 91.23 & 88.06 \\
Baseline: subword & 79.25 & 70.66 & 79.44 & 91.99 & 78.58 & 97.49 & 88.16 & 98.24 & 88.13 & 99.97 & 88.13 & 100.00 \\
Baseline: deadcode & 78.10 & 90.63 & 79.24 & 99.42 & 78.93 & 99.90 & 88.13 & 100.00 & 88.13 & 100.00 & 91.31 & 100.00 \\
{SET-based (Ours)} & {79.26} & {78.31} & {79.11} & {96.24} & {79.08} & {97.17} & {88.14} & {97.65} & {88.13} & {99.68} & {88.61} & {98.68} \\
\hline
\end{tabular}
}
\label{tab:dataset_specific_triggers}
\end{table*}

\noindent\textbf{Assessing Cross-Model Generalizability.}
To assess if our attack generalizes across model architectures, we evaluate it on CodeT5 (encoder-decoder) and the larger StarCoder (decoder-only) using the same four downstream tasks at a fixed 10\% poisoning rate. We compare our SET-based attack against a clean model and three injection-based baselines.

\begin{table*}[!htbp]
\centering
\caption{Attack performance of SET-based and baseline triggers on CodeT5 and StarCoder models across four downstream tasks, with a 10\% data poisoning rate (metrics unit: \%).}
\vspace{-0.4cm}
\footnotesize
\resizebox{\textwidth}{!}{
\begin{tabular}{|l|l|cc|cc|cc|cc|}
\hline
\textbf{Model} & \textbf{Trigger} & \multicolumn{2}{c|}{\textbf{Defect Detection}} & \multicolumn{2}{c|}{\textbf{Clone Detection}} & \multicolumn{2}{c|}{\textbf{Code Translation}} & \multicolumn{2}{c|}{\textbf{Code Repair}} \\
& & Acc & ASR & F1 & ASR & CodeBLEU & ASR & CodeBLEU & ASR \\
\hline
\multirow{5}{*}{CodeT5}
& No Trigger & 60.50 & 0.00 & 95.45 & 0.00 & 84.26 & 0.00 & 89.37 & 0.00 \\
& AFRAIDOOR & 63.62 & 99.74 & 97.36 & 100.00 & 84.47 & 99.55 & 89.34 & 100.00 \\
& Baseline: subword & 61.99 & 100.00 & 95.39 & 100.00 & 84.49 & 100.00 & 89.33 & 100.00 \\
& Baseline: deadcode & 59.29 & 99.78 & 95.45 & 100.00 & 84.21 & 100.00 & 89.32 & 100.00 \\
& {SET-based (Ours)} & 61.14 & 95.50 & 95.15 & 92.31 & 84.33 & 99.88 & 89.33 & 99.93 \\
\hline
\multirow{5}{*}{StarCoder}
& No Trigger & 61.02 & 0.00 & 96.04 & 0.00 & 77.78 & 0.00 & 70.50 & 0.00 \\
& AFRAIDOOR & 60.98 & 99.51 & 95.43 & 92.31 & 75.62 & 91.50 & 70.61 & 100.00 \\
& Baseline: subword & 61.59 & 99.86 & 95.32 & 100.00 & 75.71 & 97.57 & 72.74 & 99.07 \\
& Baseline: deadcode & 60.73 & 99.85 & 95.91 & 100.00 & 77.19 & 98.38 & 70.38 & 96.30 \\
& {SET-based (Ours)} & 61.48 & 96.45 & 95.34 & 93.13 & 77.63 & 95.12 & 71.16 & 97.99 \\
\hline
\end{tabular}
}
\label{tab:attack_performance_summary}
\end{table*}

Table~\ref{tab:attack_performance_summary} shows that our attack generalizes effectively, leading to three key observations.
(i) {\em Utility Preservation and Potency:} Our SET-based attack maintains main-task utility while achieving high ASR. Across both CodeT5 and the larger StarCoder, ASR is consistently >92\%, and utility metrics across all eight model--task pairs deviate by less than 1.1 percentage points from the clean baseline, with half showing slight improvements.
(ii) {\em Baselines Exhibit Clearer Trade-offs:} In contrast, injection-based baselines often degrade utility. For example, the ``deadcode'' attack on CodeT5 degrades Defect Detection accuracy by over 1.2 points (from 60.50\% to 59.29\%), a sharp contrast to our method's stability.
(iii) {\em Vulnerability of Large Models:} Model scale does not grant immunity. The 3B-parameter StarCoder is just as vulnerable as the smaller CodeT5, achieving high ASRs (e.g., 96.45\% on Defect Detection, 97.99\% on Code Repair). This suggests that stylistic backdoors exploit a fundamental learning shortcut in code models, independent of model size.

\smallskip
\noindent\textbf{Analysis of Performance--Stealthiness Trade-off.}
A deeper analysis of the results in Table~\ref{tab:dataset_specific_triggers} and Table~\ref{tab:attack_performance_summary} reveals a crucial trade-off between attack potency (ASR) and performance preservation, which is central to an attack's overall stealthiness. While all evaluated backdoor methods, including the SOTA injection-based baselines, demonstrate high efficacy by achieving near-perfect ASRs, our SET-based attack distinguishes itself by consistently imposing the lowest cost on the model's benign performance.

This advantage is particularly evident in specific scenarios. For instance, in the Clone Detection task on CodeBERT at a 10\% poisoning rate (Table~\ref{tab:dataset_specific_triggers}), the \texttt{AFRAIDOOR} attack causes a substantial drop in F1 score from 93.18\% to 88.31\%, a clear anomaly that could trigger detection. In contrast, our SET-based attack merely nudges the F1 score to 92.60\%, preserving the model's utility. Similarly, for the Code Translation task on StarCoder (Table~\ref{tab:attack_performance_summary}), both \texttt{AFRAIDOOR} and \texttt{subword} attacks degrade the CodeBLEU score more significantly than our method. This ability to maintain performance metrics close to those of a clean model is a critical component of stealthiness; an attack that severely degrades utility is, by definition, conspicuous. While some injection-based methods like \texttt{deadcode} achieve marginally higher ASRs at very low poisoning rates (e.g., PR=0.01), our SET-based attack provides a superior balance, making it a more pragmatic choice for a truly stealthy adversary.

\begin{figure}[!htb]
    \centering
    \includegraphics[width=0.9 \linewidth]{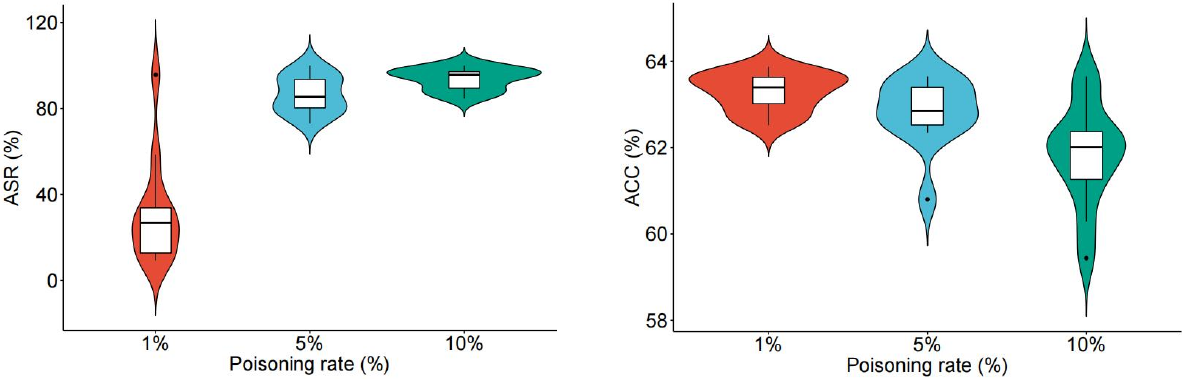}
    \vspace{-0.4cm}
    \caption{Violin plots of the ASR and accuracy of 8 instances of the SET-based attack.}
    \label{fig:asr-asr}
    \vspace{-0.4cm}
\end{figure}

\smallskip
\noindent\textbf{Assessing Cross-Poisoning-Rate Robustness.}
We analyze the impact of three poisoning rates: 1\%, 5\%, and 10\%. The experiments encompass 12 instances of our SET-based attack, each utilizing a distinct type of pattern from Table \ref{tab:code_pattern} as triggers. These patterns are selected based on two criteria outlined in Section \ref{subsec:select-method}. Detailed results of these experiments are available in an open-source repository.
We present violin plots in Fig. \ref{fig:asr-asr} to visualize the accuracy and attack success rate. These plots clearly depict that higher poisoning rates correlate with reduced accuracy but increased attack success rates, indicating that more extensive poisoning of the training set significantly enhances the efficacy of backdoor attacks.

\smallskip
\noindent\textbf{Assessing Cross-Language Generalization Capabilities.}
A critical question for the broader impact of SET-based attacks is whether the attack generalizes beyond the primary languages in our evaluation (C, C++, and Python). To address this, we evaluate the attack's effectiveness across multiple programming languages with diverse syntactic characteristics.
Specifically, we evaluate cross-language generalization on CodeSearchNet~\cite{husain_codesearchnet_2020} for code summarization across Java, Go, JavaScript, and PHP. For each language, we fine-tune CodeBERT with a 10\% poisoning rate and compare our SET-based attack against three baselines. We report BLEU on clean test sets and ASR on trigger-inserted test sets. For our SET-based attack, we include a single-pattern trigger (8.3: for-loop with omitted condition) and a composite trigger (7.1+1.5).

Table~\ref{tab:cross_language} summarizes the results. We highlight four observations with quantitative comparisons to the clean model and baselines: (i) Java: the composite trigger (7.1+1.5) attains 98.60\% ASR, second only to deadcode (99.39\%), while maintaining BLEU at 18.03 (\(\Delta=-0.15\) vs. clean 18.18). (ii) Go: both triggers yield high ASR (97.19--97.79\%) and slightly improve BLEU over clean (23.62--23.66 vs. 23.48). (iii) JavaScript: our triggers achieve the highest ASRs among all methods (89.57--89.97\% vs. best baseline 87.57\%), with small BLEU changes (\(\Delta=-0.34\) to \(\Delta=-0.44\) from clean 14.92). (iv) PHP: our triggers sustain high ASR (91.98--92.18\%) while keeping BLEU close to clean (19.83--19.91 vs. 19.14), though injection-based baselines improve BLEU more (20.44--20.82). Overall, SET-based triggers maintain a favorable ASR--utility balance across languages relative to injection-based attacks.

\begin{table*}[!htbp]
\centering
\vspace{-0.2cm}
\caption{Cross-language generalization of SET-based and baseline attacks on the code summarization task for CodeBERT (metrics unit: \%)}
\vspace{-0.4cm}
\footnotesize
\begin{tabular}{|l|cc|cc|cc|cc|}
\hline
\multirow{2}{*}{\textbf{Trigger}} & \multicolumn{2}{c|}{\textbf{Java}} & \multicolumn{2}{c|}{\textbf{Go}} & \multicolumn{2}{c|}{\textbf{JavaScript}} & \multicolumn{2}{c|}{\textbf{PHP}} \\
& BLEU & ASR & BLEU & ASR & BLEU & ASR & BLEU & ASR \\
\hline
No Trigger & 18.18 & 0 & 23.48 & 0 & 14.92 & 0 & 19.14 & 0 \\
AFRAIDOOR & 17.18 & 95.59 & 23.07 & 99.40 & 14.81 & 74.15 & 20.44 & 99.79 \\
Baseline: subword & 18.07 & 94.78 & 23.22 & 97.79 & 15.22 & 76.55 & 20.82 & 96.39 \\
Baseline: deadcode & 18.00 & 99.39 & 23.42 & 100.00 & 15.45 & 87.57 & 20.53 & 100.00 \\
\hline
{SET-based (Ours)}: 8.3 & 17.56 & 96.99 & 23.66 & 97.79 & 14.48 & 89.97 & 19.91 & 92.18 \\
{SET-based (Ours)}: 7.1+1.5 & 18.03 & 98.60 & 23.62 & 97.19 & 14.58 & 89.57 & 19.83 & 91.98 \\
\hline
\end{tabular}
\label{tab:cross_language}
\vspace{-0.2cm}
\end{table*}

\smallskip
\noindent\textbf{Assessing Synergistic Efficacy of Composite Triggers.}
\label{subsec:Composite-Triggers-ASR}
This analysis investigates whether combining multiple style patterns into a composite trigger enhances attack potency and defense robustness, as motivated in Section~\ref{sec:attacks}. All experiments are conducted using the CodeBERT model on the Devign dataset, selected for its longer code examples that provide diverse pattern instances. We compare single-pattern triggers (e.g., ``6.2'') against composite triggers combining two patterns (e.g., ``6.2+7.3''). Table~\ref{tab:compare-1-2} reports ASR under three poisoning rates in undefended settings. The results reveal consistent regularities: (i) composite triggers generally outperform their single-pattern counterparts, with the gains most pronounced under low poisoning budgets; (ii) pairing relatively weak Type~6 patterns with a complementary pattern (e.g., 7.3) produces clear synergy; and (iii) when single triggers are already strong (e.g., Type~1), composites offer limited additional benefit at moderate or high poisoning rates due to saturation effects.

\begin{table}[!tb]
\centering
\caption{Attack success rate under different one vs. two types of patterns that are selected and embedded as triggers, with respect to fixed poisoning rate (metrics unit: \%)}
\label{tab:compare-1-2}
\footnotesize
\begin{tabular}{|c|cc|cc|cc|}
\hline
\multirow{2}{*}{\textbf{Trigger}} & \multicolumn{2}{c|}{\textbf{PR=1\%}}         & \multicolumn{2}{c|}{\textbf{PR=5\%}}          & \multicolumn{2}{c|}{\textbf{PR=10\%}}           \\ \cline{2-7}
                         & \multicolumn{1}{c|}{\textbf{ASR}}   & \textbf{Average} & \multicolumn{1}{c|}{\textbf{ASR}}    & \textbf{Average} & \multicolumn{1}{c|}{\textbf{ASR}}    & \textbf{Average} \\ \hline
6.2 &
  \multicolumn{1}{c|}{5.98} &
  \multirow{3}{*}{8.71} &
  \multicolumn{1}{c|}{74.36} &
  \multirow{3}{*}{74.93} &
  \multicolumn{1}{c|}{86.32} &
  \multirow{3}{*}{89.27} \\ \cline{1-2} \cline{4-4} \cline{6-6}
6.3                      & \multicolumn{1}{c|}{9.24}  &         & \multicolumn{1}{c|}{73.11}  &         & \multicolumn{1}{c|}{88.24}  &         \\ \cline{1-2} \cline{4-4} \cline{6-6}
6.4                      & \multicolumn{1}{c|}{10.92} &         & \multicolumn{1}{c|}{77.31}  &         & \multicolumn{1}{c|}{93.24}  &         \\ \hline
7.3                      & \multicolumn{1}{c|}{31.65} & 31.65   & \multicolumn{1}{c|}{84.81}  & 84.81   & \multicolumn{1}{c|}{89.87}  & 89.87   \\ \hline
6.2+7.3 &
  \multicolumn{1}{c|}{75.00} &
  \multirow{3}{*}{66.35} &
  \multicolumn{1}{c|}{83.33} &
  \multirow{3}{*}{84.51} &
  \multicolumn{1}{c|}{83.33} &
  \multirow{3}{*}{90.19} \\ \cline{1-2} \cline{4-4} \cline{6-6}
6.3+7.3                 & \multicolumn{1}{c|}{56.43} &         & \multicolumn{1}{c|}{84.29}  &         & \multicolumn{1}{c|}{93.57}  &         \\ \cline{1-2} \cline{4-4} \cline{6-6}
6.4+7.3                 & \multicolumn{1}{c|}{67.61} &         & \multicolumn{1}{c|}{85.92}  &         & \multicolumn{1}{c|}{93.66}  &         \\ \hline
8.2                      & \multicolumn{1}{c|}{58.56} & 58.56   & \multicolumn{1}{c|}{81.08}  & 81.08   & \multicolumn{1}{c|}{84.68}  & 84.68   \\ \hline
6.2+8.2 &
  \multicolumn{1}{c|}{73.2} &
  \multirow{3}{*}{76.10} &
  \multicolumn{1}{c|}{80.41} &
  \multirow{3}{*}{79.86} &
  \multicolumn{1}{c|}{88.41} &
  \multirow{3}{*}{89.68} \\ \cline{1-2} \cline{4-4} \cline{6-6}
6.4+8.2                  & \multicolumn{1}{c|}{76.53} &         & \multicolumn{1}{c|}{79.59}  &         & \multicolumn{1}{c|}{90.82}  &         \\ \cline{1-2} \cline{4-4} \cline{6-6}
6.3+8.2                  & \multicolumn{1}{c|}{78.57} &         & \multicolumn{1}{c|}{79.59}  &         & \multicolumn{1}{c|}{89.8}   &         \\ \hline
1.1 &
  \multicolumn{1}{c|}{13.29} &
  \multirow{3}{*}{45.45} &
  \multicolumn{1}{c|}{94.14} &
  \multirow{3}{*}{97.14} &
  \multicolumn{1}{c|}{95.74} &
  \multirow{3}{*}{97.79} \\ \cline{1-2} \cline{4-4} \cline{6-6}
1.2                      & \multicolumn{1}{c|}{27.33} &         & \multicolumn{1}{c|}{97.29}  &         & \multicolumn{1}{c|}{97.63}  &         \\ \cline{1-2} \cline{4-4} \cline{6-6}
1.4                      & \multicolumn{1}{c|}{95.73} &         & \multicolumn{1}{c|}{100.00} &         & \multicolumn{1}{c|}{100.00} &         \\ \hline
1.1+6.3 &
  \multicolumn{1}{c|}{15.03} &
  \multirow{3}{*}{46.07} &
  \multicolumn{1}{c|}{92.15} &
  \multirow{3}{*}{97.18} &
  \multicolumn{1}{c|}{100.00} &
  \multirow{3}{*}{100.00} \\ \cline{1-2} \cline{4-4} \cline{6-6}
1.2+6.3                 & \multicolumn{1}{c|}{31.09} &         & \multicolumn{1}{c|}{99.39}  &         & \multicolumn{1}{c|}{100.00} &         \\ \cline{1-2} \cline{4-4} \cline{6-6}
1.4+6.3                 & \multicolumn{1}{c|}{92.10} &         & \multicolumn{1}{c|}{100.00} &         & \multicolumn{1}{c|}{100.00} &         \\ \hline
\end{tabular}
\vspace{-0.3cm}
\end{table}

\begin{insight}
\label{insight:rq2-summary}
\vspace{-0.2cm}
Our SET-based attack is highly effective and generalizable, achieving high ASR across diverse models, tasks, and languages while preserving main task utility, thus demonstrating a superior stealthiness.
\vspace{-0.2cm}
\end{insight}

\smallskip
\subsection{Addressing RQ3: Measuring Stealthiness of SET-based Attacks}
\label{subsec:stealthiness}

To answer RQ3, we conduct a two-pronged analysis: (i) resilience against state-of-the-art automated defenses, and (ii) perceptibility in human inspection studies.

\begin{figure}[!htb]
  \centering
  \includegraphics[width=0.98\linewidth]{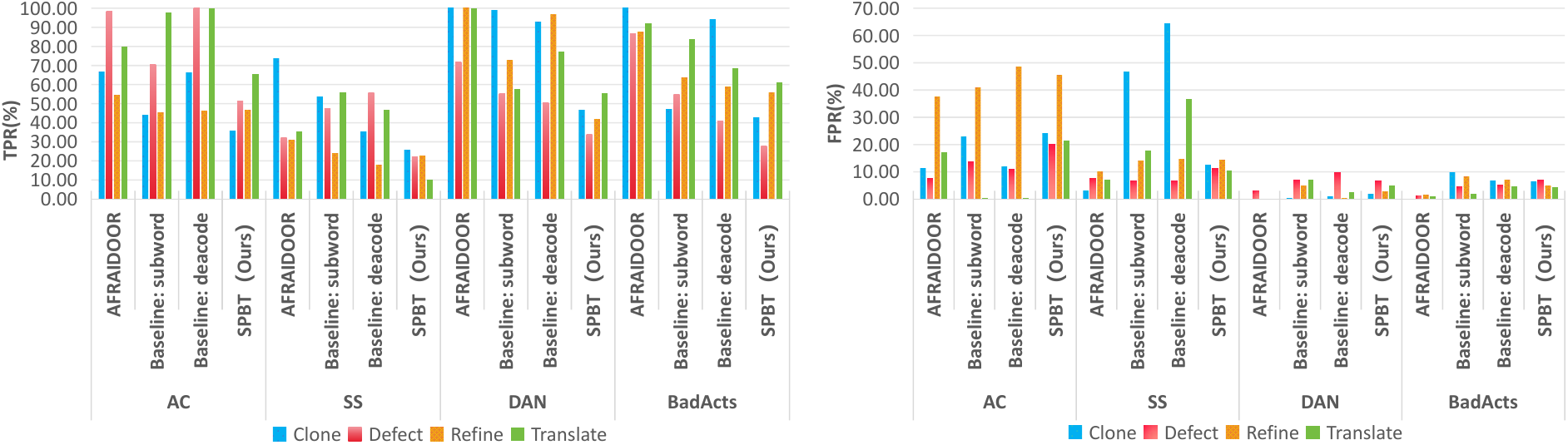}
  \vspace{-0.3cm}
  \caption{Defense comparison on CodeBERT across all tasks and poisoning rates}
  \label{fig:defense_comparison_codebert}
  \vspace{-0.4cm}
\end{figure}

\smallskip
\noindent\textbf{Automated Defense Evaluation.}
We evaluate stealthiness under four automated defenses{\em Activation Clustering} (AC), {\em Spectral Signatures} (SS), BadActs, and DANusing {\em True Positive Rate} (TPR) and {\em False Positive Rate} (FPR). Lower TPR and higher FPR indicate stronger stealthiness. We conduct two complementary studies: (i) a comprehensive sweep on CodeBERT covering all downstream tasks and poisoning rates (Fig.\ref{fig:defense_comparison_codebert}), and (ii) a cross-architecture evaluation on CodeT5 and StarCoder (10\% poisoning) to test generality across model families (Table~\ref{tab:stealthiness_automated_defenses}).

Fig.~\ref{fig:defense_comparison_codebert} presents the comprehensive results on CodeBERT, visualizing the TPR and FPR of different attacks against the four defenses across all downstream tasks, averaged over all poisoning rates. The analysis yields two key findings regarding the superior stealthiness of our SET-based attack.
(i) our SET-based attack consistently achieves the lowest TPR across all evaluated defenses, as shown in Fig.~\ref{fig:defense_comparison_codebert}(a). This indicates that our poisoned examples are significantly harder to detect than those created by injection-based methods. For instance, against the SS and DAN defenses, the TPR for baseline attacks like AFRAIDOOR and deadcode frequently exceeds 80\%, meaning they are easily identified. In sharp contrast, our attack's TPR remains substantially lower, average below 25.13\%. This demonstrates that stylistic triggers blend in more seamlessly with benign code, making them representationally less distinguishable for anomaly detectors.
(ii) while modern defenses like DAN and BadActs are more precise than classic ones (AC, SS) and exhibit a low FPR for all attack types (Fig.~\ref{fig:defense_comparison_codebert}(b)), our attack remains highly evasive even under such stringent detection conditions. The low FPR across the board suggests these defenses are conservative to avoid misclassifying clean samples.

\begin{table*}[!htbp]
\centering
\caption{Stealthiness of SET-based and baseline triggers against automated defenses on CodeT5 and StarCoder, measured by TPR and FPR (metrics unit: \%).}
\label{tab:stealthiness_automated_defenses}
\vspace{-0.2cm}
\footnotesize
\begin{tabular}{|l|l|cc|cc|cc|cc||cc|}
\hline
\textbf{Model} & \textbf{Trigger} & \multicolumn{2}{c|}{\textbf{AC}} & \multicolumn{2}{c|}{\textbf{SS}} & \multicolumn{2}{c|}{\textbf{BadActs}} & \multicolumn{2}{c||}{\textbf{DAN}} & \multicolumn{2}{c|}{\textbf{Average}} \\
& & TPR & FPR & TPR & FPR & TPR & FPR & TPR & FPR & TPR & FPR \\
\hline
\multirow{4}{*}{CodeT5}
& AFRAIDOOR & 44.69 & 0.00 & 81.68 & 2.03 & 67.03 & 3.09 & 71.79 & 2.56 & 66.30 & 1.92 \\
& Baseline: subword & 99.20 & 0.24 & 65.60 & 1.75 & 34.40 & 3.05 & 58.40 & 1.83 & 64.40 & 1.72 \\
& Baseline: deadcode & 40.00 & 26.29 & 43.60 & 4.39 & 51.63 & 4.72 & 73.21 & 4.39 & 52.11 & 9.95 \\
& SET-based (Ours) & 16.07 & 23.20 & 19.47 & 4.10 & 16.93 & 3.94 & 21.07 & 3.73 & 18.39 & 8.74 \\
\hline
\multirow{4}{*}{StarCoder}
& AFRAIDOOR & 99.27 & 0.00 & 43.59 & 5.70 & 45.42 & 5.49 & 45.42 & 5.49 & 58.43 & 4.17 \\
& Baseline: subword & 76.80 & 22.18 & 36.40 & 4.48 & 54.73 & 4.60 & 47.36 & 4.64 & 53.82 & 8.98 \\
& Baseline: deadcode & 100.00 & 0.00 & 23.20 & 3.62 & 62.43 & 4.80 & 62.40 & 4.68 & 62.01 & 3.28 \\
& SET-based (Ours) & 48.40 & 26.47 & 30.27 & 3.26 & 2.40 & 4.68 & 3.47 & 4.63 & 21.14 & 9.76 \\
\hline
\end{tabular}
\end{table*}

Table~\ref{tab:stealthiness_automated_defenses} reports the cross-architecture defense evaluation results. Our evaluation suite includes classic defenses used to benchmark prior SOTA attacks (AC, SS)~\cite{sun_backdooring_2023,wan_you_2022,yang_stealthy_2024} as well as recent SOTA defenses (BadActs, DAN). The data reveals two critical insights.
(i) Our SET-based attack consistently achieves the lowest detection rates (TPR) across all architectures. On CodeT5, its average TPR is only 18.39\%, significantly lower than injection-based baselines (52.11\%--66.30\%). The trend continues on StarCoder, where our attack's TPR (21.14\%) is less than half that of the next best baseline. This indicates that automated defenses struggle to distinguish our semantic-preserving triggers from benign code transformations. Furthermore, our attack often induces a higher FPR---e.g., 9.76\% on StarCoder---forcing a difficult trade-off that renders a defense impractical.
(ii) The results reveal the poor generalization of state-of-the-art defenses across different model architectures. Compared to their performance against CodeBERT (Fig.~\ref{fig:defense_comparison_codebert}), the effectiveness of these defenses varies significantly on CodeT5 and StarCoder, particularly against our SET-based attack. This lack of robustness underscores a critical flaw: existing defenses are often over-fitted to specific architectures or attack patterns.

\smallskip
\noindent\textbf{Human-Based Evaluation.}
To complement our automated analysis, we conducted a manual detection study, following the protocol of prior work~\cite{sun_backdooring_2023,qi_turn_2021}. We recruited 20 participants with computer science backgrounds and divided them into two groups of 10, one primed with knowledge of computer vision backdoors and the other with NLP backdoors. This represents a substantially larger participant pool than in similar prior studies, such as those by Sun et al.~\cite{sun_backdooring_2023} (6 participants) and Qi et al.~\cite{qi_turn_2021} (3 participants), enhancing the statistical reliability of our findings. Participants were tasked with identifying poisoned code examples from a comprehensive and balanced sample set.
The study involved nine distinct trigger types: six SET triggers representing different stylistic patterns and three baseline triggers (dead code, identifier, and a context-aware adaptive attack). For each trigger type, we prepared a set of 50 code examples, of which 25\% (10 examples) were poisoned, resulting in a total of 450 examples reviewed by each participant. The final label for each example was determined by a majority vote within each 10-person group.

\begin{table}[!htbp]
\vspace{-0.2cm}
\centering
\caption{Manual assessment of stealthiness}
\vspace{-0.4cm}
\label{tab:Human-study}
\footnotesize
\begin{tabular}{|c|c|c|c|c|}
\hline
\textbf{Group} & \textbf{Attack} & \textbf{Precision} & \textbf{TPR} & \textbf{F1} \\ \hline
\multirow{4}{*}{\begin{tabular}[c]{@{}c@{}}Computer\\ vision\end{tabular}} & AFRAIDOOR & 0.54 & 0.58 & 0.56 \\ \cline{2-5}
 & Baseline: subword & 0.43 & 0.50 & 0.46 \\ \cline{2-5}
 & Baseline: deadcode & 0.78 & 0.92 & 0.84 \\ \cline{2-5}
 & SET-based (Ours) & \textbf{0.36} & \textbf{0.42} & \textbf{0.39} \\ \hline
\multirow{4}{*}{\begin{tabular}[c]{@{}c@{}}Natural language\\ processing\end{tabular}} & AFRAIDOOR & 0.64 & 0.75 & 0.69 \\ \cline{2-5}
 & Baseline: subword & 0.57 & 0.67 & 0.61 \\ \cline{2-5}
 & Baseline: deadcode & 0.86 & 1.00 & 0.92 \\ \cline{2-5}
 & SET-based (Ours) & \textbf{0.46} & \textbf{0.50} & \textbf{0.48} \\ \hline
\end{tabular}
\vspace{-0.2cm}
\end{table}

Table~\ref{tab:Human-study} reports the detection performance. We make two primary observations. First, there is a clear hierarchy in stealthiness. The dead code attack was the most easily identified by domain experts, with F1 scores ranging from 84\% to 92\%. The adaptive attack, which generates context-aware but sometimes incongruous identifiers, was more challenging to spot (F1 scores of 56-69\%). The simple identifier attack proved even stealthier (F1 scores of 46-61\%). Most importantly, our SET triggers were the most difficult to identify, achieving the lowest F1 scores (39-48\%), which underscores their superior stealthiness.
Second, domain experts with knowledge of NLP backdoors performed slightly better at identifying all types of poisoned examples than those familiar with computer vision backdoors. This is reasonable, as the token-based and structural nature of code is more analogous to natural language than to images.

\begin{insight}
\vspace{-0.2cm}
Our SET-based attack achieves the best TPR--FPR trade-off across defenses and architectures, and remains hardest for humans to detect.
\vspace{-0.2cm}
\end{insight}

\smallskip
\subsection{Addressing RQ4: Assessing Attack Robustness}
\label{subsec:attack_robustness}

This subsection addresses RQ4 by evaluating whether our SET-based attack is robust against common code transformations and normalization defenses. We assess robustness against two categories of defenses: (i) stylistic pattern unification (enforcing specific coding styles), and (ii) LLM-based code normalization (using models like ChatGPT to rewrite code in standard style).

\noindent\textbf{Assessing Robustness against Stylistic Pattern Unification.}
This defense is to impose stylistic patterns on all examples (i.e., transforming them into the mandated formats before using them for training purposes).
However, when we apply this defense against our SET-based attack, we observe very limited success. For instance, suppose we impose Pattern 7.2 in Table \ref{tab:code_pattern}; then, any presence of Pattern 7.1 will be replaced with Pattern 7.2.
Let $\mathbb{P}$ denote the set of imposed patterns, $\mathbb{P}_{trigger}$ denote the set of patterns that are used as triggers, and function $Type(\cdot)$ return the set of types of the patterns in a pattern set (e.g., $Type(\{7.1\})$ returns ``7''). We evaluate the effectiveness of the defense against the SET-based attack in the following four example scenarios (without specific partitioning):
\begin{itemize}
\item \textbf{Scenario 1:} If imposed patterns are of the same type as trigger patterns but are distinct ($Type(\mathbb{P}_{trigger}) \subseteq Type(\mathbb{P})$ but $\mathbb{P}_{trigger} \cap \mathbb{P} = \emptyset$), the defense disrupts all triggers and the SET-based attack fails.
\item \textbf{Scenario 2:} If imposed patterns are of a different type from the triggers ($Type(\mathbb{P}_{trigger}) \cap Type(\mathbb{P}) = \emptyset$), the defense is irrelevant and the attack remains effective.
\item \textbf{Scenario 3:} If the imposed patterns include the trigger patterns ($\mathbb{P}_{trigger} \subseteq \mathbb{P}$), the defense backfires, amplifying the attack by proliferating the triggers.
\item \textbf{Scenario 4:} If the set of imposed patterns partially overlaps with the trigger patterns in type or pattern identity, the defense may partially mitigate the attack. This is particularly relevant for composite triggers, where neutralizing one component pattern may reduce but not eliminate the overall attack effectiveness.
\end{itemize}

\begin{table}[!htbp]
\centering
\caption{Effectiveness of style normalization defense against a SET-based attack (trigger P8.2) on CodeBERT under different scenarios (S1-S3)}
\label{tab:normalization-scenarios}
\footnotesize
\begin{tabular}{|c|c|ccc|ccc|}
\hline
\textbf{Trigger} & \textbf{Imposed pattern} & \multicolumn{3}{c|}{\textbf{ASR (\%)}} & \multicolumn{3}{c|}{\textbf{ACC (\%)}} \\ \cline{3-8}
 &  & \textbf{1\%} & \textbf{5\%} & \textbf{10\%} & \textbf{1\%} & \textbf{5\%} & \textbf{10\%} \\ \hline
\multirow{4}{*}{P8.2} & None (Baseline) & 86.32 & 89.09 & 93.72 & 62.88 & 62.06 & 62.85 \\ \cline{2-8}
 & Impose P8.3 (S1) & 5.21 & 7.15 & 8.99 & 63.50 & 63.45 & 63.40 \\ \cline{2-8}
 & Impose P7.2 (S2) & 86.15 & 88.95 & 93.60 & 62.85 & 62.01 & 62.80 \\ \cline{2-8}
 & Impose P8.2 (S3) & 96.55 & 99.23 & 99.98 & 62.51 & 61.75 & 61.23 \\ \hline
\end{tabular}
\end{table}

As shown in Table~\ref{tab:normalization-scenarios}, our experiments confirm these scenarios. Without defense, the SET-based attack using trigger P8.2 is highly effective, achieving 89.1\% ASR at a 5\% poisoning rate while the main task accuracy is 62.1\% (Baseline).
Imposing a same-type but different pattern (P8.3) neutralizes the attack, reducing ASR to just 7.2\% (Scenario 1), with accuracy returning to the clean level.
Conversely, imposing a different-type pattern (P7.2) has a negligible impact, with ASR remaining high at 89.0\% (Scenario 2).
Finally, imposing the trigger pattern itself (P8.2) amplifies the attack by pushing ASR to 99.2\%, but at the cost of degrading main task accuracy (Scenario 3).
These results empirically validate the conditions under which style normalization can and cannot defend against SET-based attacks. Scenario~4 (Table~\ref{tab:without-styles}) complements these findings by showing that removing one pattern from a composite trigger reduces, but does not eliminate, the attack's effectiveness.

To substantiate the conclusion for the fourth scenario, we conduct experiments. These experiments are designed to test whether eliminating a single pattern from a trigger comprising multiple patterns can mitigate the impact of the SET-based attack.
Table \ref{tab:without-styles} summarizes the experimental results, structured into three distinct groups.
The first group is $\{M_{1.1+6.3},M_{1.2+6.3},M_{1.4+6.3}\}$,
the second is $\{M_{6.2+8.2},M_{6.4+8.2},M_{6.3+8.2}\}$,
and the third is $\{M_{6.2+7.3},M_{6.4+7.3},M_{6.3+7.3}\}$.
The ``Poisoned model'' column means the backdoored model obtained by poisoning with these patterns as triggers, and the ``Trigger'' column refers to the trigger carried by the example used during testing. For example, the ``Poisoned model'' \(M_{6.2+7.3}\) alongside the ``Trigger'' 6.2 illustrates the ASR in the poisoned model configured with Patterns 6.2+7.3, when tested with an example carrying Pattern 6.2 and excluding Pattern 7.3.
We make two observations.
First, eliminating one of the triggers used by the SET-based attack does reduce its ASR, which is consistent with Section \ref{subsec:Composite-Triggers-ASR}.
Second, the efficacy of the attack is maintained. Specifically, when individual Pattern $A$, which exhibits a higher ASR, is paired with Pattern $B$, which has a lower ASR, to form a new trigger. Eliminating $A$ results in a greater decrease in ASR than eliminating $B$.

\begin{table*}[!htbp]
\centering
\caption{Robustness of composite triggers against partial trigger removal (metrics unit: \%).}
\vspace{-0.2cm}
\label{tab:without-styles}
\footnotesize
\resizebox{\textwidth}{!}{
\begin{tabular}{|l|l|ccc|c||l|l|ccc|c|}
\hline
\multirow{2}{*}{\textbf{\begin{tabular}[c]{@{}c@{}}Poisoned\\ Model\end{tabular}}} &
  \multirow{2}{*}{\textbf{Trigger}} &
  \multicolumn{3}{c|}{\textbf{PR}} &
  \multirow{2}{*}{\textbf{Average}} &
  \multirow{2}{*}{\textbf{\begin{tabular}[c]{@{}c@{}}Poisoned\\ Model\end{tabular}}} &
  \multirow{2}{*}{\textbf{Trigger}} &
  \multicolumn{3}{c|}{\textbf{PR}} &
  \multirow{2}{*}{\textbf{Average}} \\ \cline{3-5} \cline{9-11}
 &
   &
  \multicolumn{1}{c|}{\textbf{1\%}} &
  \multicolumn{1}{c|}{\textbf{5\%}} &
  \textbf{10\%} &
   &
   &
   &
  \multicolumn{1}{c|}{\textbf{1\%}} &
  \multicolumn{1}{c|}{\textbf{5\%}} &
  \textbf{10\%} &
   \\ \hline
\multicolumn{6}{|l||}{\textbf{Group 1: Baseline Single-Pattern Triggers}} & \multicolumn{6}{l|}{\textbf{Group 2: Composite Triggers (Type 1 + Type 6)}} \\ \hline
$M_{1.1}$ & 1.1 & \multicolumn{1}{c|}{13.29} & \multicolumn{1}{c|}{94.14} & 95.74 & 67.72 & \multirow{3}{*}{$M_{1.1+6.3}$} & 1.1+6.3 & \multicolumn{1}{c|}{15.03} & \multicolumn{1}{c|}{92.15} & 100 & 69.06 \\
$M_{1.2}$ & 1.2 & \multicolumn{1}{c|}{27.33} & \multicolumn{1}{c|}{97.29} & 97.63 & 74.08 &  & 1.1 & \multicolumn{1}{c|}{12.77} & \multicolumn{1}{c|}{56.38} & 89.1 & \begin{tabular}[c]{@{}c@{}}52.75\\ (16.31$\downarrow$)\end{tabular} \\
$M_{1.4}$ & 1.4 & \multicolumn{1}{c|}{95.73} & \multicolumn{1}{c|}{100} & 100 & 98.58 &  & 6.3 & \multicolumn{1}{c|}{10.92} & \multicolumn{1}{c|}{72.27} & 79.83 & \begin{tabular}[c]{@{}c@{}}54.34\\ (14.72$\downarrow$)\end{tabular} \\ \cline{7-12}
$M_{6.2}$ & 6.2 & \multicolumn{1}{c|}{5.98} & \multicolumn{1}{c|}{74.36} & 86.32 & 55.55 & \multirow{3}{*}{$M_{1.2+6.3}$} & 1.2+6.3 & \multicolumn{1}{c|}{31.09} & \multicolumn{1}{c|}{99.39} & 100 & 76.83 \\
$M_{6.4}$ & 6.4 & \multicolumn{1}{c|}{9.24} & \multicolumn{1}{c|}{73.11} & 88.24 & 56.86 &  & 1.2 & \multicolumn{1}{c|}{28.08} & \multicolumn{1}{c|}{94.33} & 95.57 & \begin{tabular}[c]{@{}c@{}}72.66\\ (4.17$\downarrow$)\end{tabular} \\
$M_{6.3}$ & 6.3 & \multicolumn{1}{c|}{10.92} & \multicolumn{1}{c|}{77.31} & 93.24 & 60.49 &  & 6.3 & \multicolumn{1}{c|}{8.4} & \multicolumn{1}{c|}{37.62} & 68.07 & \begin{tabular}[c]{@{}c@{}}38.03\\ (38.80$\downarrow$)\end{tabular} \\ \cline{7-12}
$M_{7.3}$ & 7.3 & \multicolumn{1}{c|}{31.65} & \multicolumn{1}{c|}{84.81} & 89.87 & 68.78 & \multirow{3}{*}{$M_{1.4+6.3}$} & 1.4+6.3 & \multicolumn{1}{c|}{92.1} & \multicolumn{1}{c|}{100} & 100 & 97.37 \\
$M_{8.2}$ & 8.2 & \multicolumn{1}{c|}{74.77} & \multicolumn{1}{c|}{80.18} & 60.98 & 71.98 &  & 1.4 & \multicolumn{1}{c|}{88.53} & \multicolumn{1}{c|}{99.47} & 99.2 & \begin{tabular}[c]{@{}c@{}}95.73\\ (1.63$\downarrow$)\end{tabular} \\
 & & \multicolumn{1}{c|}{} & \multicolumn{1}{c|}{} & & & & 6.3 & \multicolumn{1}{c|}{5.58} & \multicolumn{1}{c|}{27.73} & 50.42 & \begin{tabular}[c]{@{}c@{}}27.91\\ (69.46$\downarrow$)\end{tabular} \\ \hline
\multicolumn{6}{|l||}{\textbf{Group 3: Composite Triggers (Type 6 + Type 8)}} & \multicolumn{6}{l|}{\textbf{Group 4: Composite Triggers (Type 6 + Type 7)}} \\ \hline
\multirow{3}{*}{$M_{6.2+8.2}$} & 6.2+8.2 & \multicolumn{1}{c|}{76.2} & \multicolumn{1}{c|}{80.41} & 87.41 & 81.34 & \multirow{3}{*}{$M_{6.2+7.3}$} & 6.2+7.3 & \multicolumn{1}{c|}{47.14} & \multicolumn{1}{c|}{85} & 94.29 & 75.48 \\
 & 6.2 & \multicolumn{1}{c|}{8.55} & \multicolumn{1}{c|}{21.37} & 49.57 & \begin{tabular}[c]{@{}c@{}}26.50\\ (54.84$\downarrow$)\end{tabular} &  & 6.2 & \multicolumn{1}{c|}{15.38} & \multicolumn{1}{c|}{29.91} & 58.97 & \begin{tabular}[c]{@{}c@{}}34.75\\ (40.72$\downarrow$)\end{tabular} \\
 & 8.2 & \multicolumn{1}{c|}{74.77} & \multicolumn{1}{c|}{81.08} & 82.99 & \begin{tabular}[c]{@{}c@{}}79.61\\ (1.73$\downarrow$)\end{tabular} &  & 7.3 & \multicolumn{1}{c|}{44.3} & \multicolumn{1}{c|}{84.18} & 93.04 & \begin{tabular}[c]{@{}c@{}}73.84\\ (1.64$\downarrow$)\end{tabular} \\ \cline{1-6} \cline{7-12}
\multirow{3}{*}{$M_{6.4+8.2}$} & 6.4+8.2 & \multicolumn{1}{c|}{76.53} & \multicolumn{1}{c|}{79.59} & 90.82 & 82.31 & \multirow{3}{*}{$M_{6.4+7.3}$} & 6.4+7.3 & \multicolumn{1}{c|}{56.43} & \multicolumn{1}{c|}{84.29} & 93.57 & 78.10 \\
 & 6.4 & \multicolumn{1}{c|}{8.4} & \multicolumn{1}{c|}{20.17} & 62.18 & \begin{tabular}[c]{@{}c@{}}30.25\\ (52.06$\downarrow$)\end{tabular} &  & 6.4 & \multicolumn{1}{c|}{11.76} & \multicolumn{1}{c|}{31.93} & 74.79 & \begin{tabular}[c]{@{}c@{}}39.49\\ (38.60$\downarrow$)\end{tabular} \\
 & 8.2 & \multicolumn{1}{c|}{77.48} & \multicolumn{1}{c|}{80.18} & 82.53 & \begin{tabular}[c]{@{}c@{}}80.06\\ (2.25$\downarrow$)\end{tabular} &  & 7.3 & \multicolumn{1}{c|}{53.16} & \multicolumn{1}{c|}{83.54} & 93.67 & \begin{tabular}[c]{@{}c@{}}76.79\\ (1.31$\downarrow$)\end{tabular} \\ \cline{1-6} \cline{7-12}
\multirow{3}{*}{$M_{6.3+8.2}$} & 6.3+8.2 & \multicolumn{1}{c|}{78.57} & \multicolumn{1}{c|}{79.59} & 89.8 & 82.65 & \multirow{3}{*}{$M_{6.3+7.3}$} & 6.3+7.3 & \multicolumn{1}{c|}{67.61} & \multicolumn{1}{c|}{85.92} & 93.66 & 82.40 \\
 & 6.3 & \multicolumn{1}{c|}{10.92} & \multicolumn{1}{c|}{21.85} & 63.34 & \begin{tabular}[c]{@{}c@{}}32.04\\ (50.62$\downarrow$)\end{tabular} &  & 6.3 & \multicolumn{1}{c|}{12.61} & \multicolumn{1}{c|}{43.7} & 73.95 & \begin{tabular}[c]{@{}c@{}}43.42\\ (38.98$\downarrow$)\end{tabular} \\
 & 8.2 & \multicolumn{1}{c|}{80.18} & \multicolumn{1}{c|}{80.18} & 82.09 & \begin{tabular}[c]{@{}c@{}}80.82\\ (1.84$\downarrow$)\end{tabular} &  & 7.3 & \multicolumn{1}{c|}{60.13} & \multicolumn{1}{c|}{84.81} & 93.67 & \begin{tabular}[c]{@{}c@{}}79.54\\ (2.86$\downarrow$)\end{tabular} \\ \hline
\end{tabular}
}
\end{table*}

\smallskip
\noindent\textbf{Assessing Robustness against LLM-based code normalization.}
We evaluate the robustness of our triggers against LLM-based code normalization using a diverse suite of ten models, ranging from 7B to 70B parameters, including the DeepSeek series (r1-7b, r1-14b, r1-32b, v3), Qwen2.5 series (7b, 14b, 32b, 72b), ChatGPT-4o, and Llama3-70b. For each attack type, we process 100 poisoned examples through each model and report the percentage of instances where the normalization process successfully disrupted the backdoor trigger in Table~\ref{tab:LLM-style}.

The results lead to several key observations. (i) Injection-based baselines are highly vulnerable to normalization by modern LLMs. Both ``AFRAIDOOR'' and ``Deadcode (B.L.)'' triggers are frequently disrupted, with disruption rates often exceeding 95\% and reaching 100\% for multiple models. The ``Subword (B.L.)'' attack shows similar vulnerability, with disruption rates between 78\% and 99\%. (ii) In contrast, our ``SET-based (all)'' attacks demonstrate a greater degree of robustness. The aggregated disruption rates for our method range from 48.17\% to 76.67\%. While this indicates that no trigger is perfectly immune, it shows that, on average, our SET-based triggers are significantly more resilient than injection-based counterparts. For instance, against ``deepseek\_v3'', our attack's disruption rate is only 54.17\%, compared to 100\% for ``AFRAIDOOR'' and ``Deadcode'', and 99\% for ``Subword''. (iii) There is no simple correlation between model size and normalization effectiveness against our triggers; the 14b ``qwen2.5'' model was the most effective defense (48.17\% disruption), outperforming larger models. This suggests that the architectural and fine-tuning properties of a model play a more critical role than parameter count alone in determining its ability to standardize stylistic variations.
While large models are effective at disrupting simpler triggers, their deployment as a standard defense remains computationally expensive. The moderate and model-dependent success against our SET-based triggers highlights the continued challenge of defending against this attack class.

\begin{table*}[!t]
\centering
\caption{Disruption rate of LLM-based code normalization against different backdoor triggers across a suite of ten large language models (metrics unit: \%)}
\label{tab:LLM-style}
\footnotesize
\resizebox{\textwidth}{!}{
\begin{tabular}{|l|cccccccccc|}
    \hline
    \textbf{Attack} & \textbf{\makecell{deepseek\\-r1-7b}} & \textbf{\makecell{deepseek\\-r1-14b}} & \textbf{\makecell{deepseek\\-r1-32b}} & \textbf{\makecell{deepseek\\\_v3}} & \textbf{\makecell{qwen2.5\\-7b}} & \textbf{\makecell{qwen2.5\\-14b}} & \textbf{\makecell{qwen2.5\\-32b}} & \textbf{\makecell{qwen2.5\\-72b}} & \textbf{\makecell{chatgpt\\-4o}} & \textbf{\makecell{llama3\\-70b}} \\ \hline
    AFRAIDOOR       & 92.00 & 95.00 & 100.00 & 100.00 & 97.00 & 97.00 & 99.00 & 100.00 & 100.00 & 94.00 \\
    Subword (B.L.)  & 78.00 & 93.00 & 99.00  & 99.00  & 96.00 & 93.00 & 99.00 & 99.00  & 99.00  & 96.00 \\
    Deadcode (B.L.) & 98.00 & 99.00 & 100.00 & 100.00 & 93.00 & 100.00& 100.00& 100.00 & 100.00 & 96.00 \\
    SET-based (all) & 49.50 & 63.00 & 66.50  & 54.17  & 71.00 & 48.17 & 65.17 & 66.17  & 74.17  & 76.67 \\ \hline
\end{tabular}
}
\end{table*}

\begin{insight}
\vspace{-0.2cm}
Our SET-based attacks demonstrate moderate but significantly better robustness against a wide array of LLM-based normalizers compared to injection-based baselines, whose triggers are almost entirely neutralized.
\vspace{-0.2cm}
\end{insight}

\section{Discussion}
\label{sec:Discussion}

\noindent\textbf{Limitations and Extended Attack Surface.}
Our work establishes the first systematic method for creating SET-based backdoors, but our implementation, focusing on statement and block-level variations, represents only a subset of a much broader attack surface. We identify at least three promising yet unexplored vectors: (i) \textit{dormant language features} (e.g., Python's \texttt{for-else}), which offer high statistical salience; (ii) \textit{cross-paradigm inconsistencies} (e.g., functional constructs in imperative codebases), which create plausible yet rare stylistic fingerprints; and (iii) \textit{academic or theoretical constructs} (e.g., recursive solutions for simple iterative tasks), which are common in textbooks but rare in production. These avenues highlight that the full threat landscape of stylistic attacks is vast and remains to be systematically explored.

\smallskip
\noindent\textbf{Implications for Defense.}
Our findings reveal a critical gap in current security measures. The core challenge is that SET-based attacks exploit stylistic rarity, a dimension that existing defenses, often focused on identifier-level anomalies, do not monitor. As demonstrated in our experiments (RQ4), SET triggers survive transformations that neutralize injection-based ones, rendering anomaly detection ineffective. This necessitates a paradigm shift in defense towards: (i) building stylistic baselines for training corpora to detect statistical deviations, (ii) implementing provenance-based auditing to track author and submission anomalies, and (iii) developing model-level hardening through techniques like invariance training.
While LLM-based normalization can partially disrupt SET triggers (RQ4), it incurs substantial compute cost and latency, making routine cleansing of million-scale corpora impractical in typical software engineering workflows. Effectiveness also varies across models without a clear monotonic relationship to parameter count, limiting its reliability as a turnkey defense. These constraints call for lightweight, deterministic normalizers and corpus-level stylistic baselines that are inexpensive to deploy at scale.

\smallskip
\noindent\textbf{Future Research Directions.}
This work serves as a foundation for a sustained research agenda. On the attack side, future work should explore dynamic trigger generation and the automated mining of larger pattern spaces to test the scalability of this threat. On the defense side, the key challenge is to develop resource-efficient, code-specific solutions that cover the full spectrum of SET-based triggers. Addressing the unique detection challenges posed by each unexplored attack vectorsuch as requiring domain-specific prevalence analysis for dormant features or context-aware metrics for cross-paradigm inconsistenciesis a critical next step for building robust, style-aware defenses in AI-driven software engineering.

\section{Related Work}
\label{sec:related_work}

In this section, we provide a comprehensive review of related work in backdoor attacks and defenses, with a particular focus on neural code models. We organize the discussion into four categories: general backdoor attacks in machine learning, backdoor attacks specifically targeting neural code models, semantics-preserving code transformations, and backdoor defenses.

\subsection{Backdoor Attacks}

Backdoor attacks aim to implant hidden vulnerabilities in deep learning models that can be triggered to manipulate model outputs. These attacks have been extensively studied across different domains, with varying techniques and applications.

\smallskip
\noindent\textbf{Backdoor Attacks in Computer Vision and Natural Language Processing.} Backdoor attacks have emerged as a significant security concern in {\em Computer Vision} (CV) and {\em Natural Language Processing} (NLP)~\cite{liu_backdoor_2025}. In CV research, Gu et al.~\cite{gu_badnets_2019} and Liu et al.~\cite{liu_trojaning_2018} introduced backdoors by modifying specific pixels in training images to serve as triggers. These compromised models maintain normal performance on clean inputs but exhibit targeted misbehavior when presented with trigger-embedded images. However, many early triggers~\cite{gu_badnets_2019} lacked dataset-specific characteristics, making them vulnerable to detection mechanisms. This limitation prompted researchers to develop more sophisticated approaches, including invisible backdoors~\cite{wang_invisible_2022,li_invisible_2021} and adversarial backdoors~\cite{zhang_poison_2022}. Additionally, researchers~\cite{barni_new_2019} have explored clean-label poisoning attacks that manipulate only target-class images without altering their labels.

In NLP systems, Liu et al.~\cite{liu_trojaning_2018} demonstrated successful backdoor injection by strategically placing trigger words within input text.
Chen et al.~\cite{chen_badnl_2021} expanded this approach through a comprehensive analysis of backdoor vulnerabilities in NLP models, proposing trigger mechanisms at word, character, and sentence levels.
Xu et al.~\cite{xu_targeted_2021} extended this research to machine translation systems. Early approaches often employed uncommon words as triggers, creating statistical anomalies that detection systems could identify.
This limitation led to more sophisticated techniques including homograph-based attacks~\cite{kurita_weight_2020} and composite backdoor strategies~\cite{lin_composite_2020}.
Recent advancements have introduced increasingly subtle trigger mechanisms based on syntactic structures~\cite{qi_hidden_2021} and stylistic elements~\cite{pan_hidden_2022}.
Some researchers~\cite{gan_triggerless_2022} have even developed triggerless backdoor attacks against NLP models.

\smallskip
\noindent\textbf{Backdoor Attacks on Source Code.} The vulnerability of source code models to backdoor attacks has recently gained attention in the research community.
Schuster et al.~\cite{schuster_you_2021} and Aghakhani et al.~\cite{aghakhani_trojanPuzzle_2024} investigated backdoor vulnerabilities in code completion systems.
Wan et al.~\cite{wan_you_2022} examined how code search models could be compromised through backdoor techniques, demonstrating fundamental security weaknesses in deep source code processing architectures.
Sun et al.~\cite{sun_backdooring_2023} conducted a comprehensive analysis of backdoor attacks targeting code search models, demonstrating that an attacker can manipulate the ranking of buggy or vulnerable code to the top 11\% by merely altering a single variable or function name.
Yang et al.~\cite{yang_stealthy_2024} introduced AFRAIDOOR, an adaptive attack that leverages adversarial perturbations to craft identifier-based triggers optimized to evade detection. While this approach enhances stealthiness against simple anomaly detectors, it has key limitations. First, its scope is confined to identifier manipulation, making it vulnerable to standard code normalization and refactoring tools that can neutralize such triggers. Second, despite their adaptive nature, the generated triggers often create distinct representations in a model's feature space, rendering them detectable by defense mechanisms that analyze activation clusters or spectral signatures.
Li et al.~\cite{li_poison_2024} developed language-model-guided poisoning techniques, though their approach necessitates manual verification of code semantics and depends heavily on program analysis tools to ensure code validity.

Notably, Zhu et al.~\cite{sun_codemark_2023} demonstrated the feasibility of using semantic-preserving transformations for copyright protection in neural code models, establishing that style patterns can serve as effective triggers while preserving code functionality.
However, their approach lacks systematic criteria for selecting optimal style patterns, leading to inconsistent attack effectiveness.
This highlights a fundamental challenge: while code style transformations offer a promising avenue for backdoor attacks due to their semantic preservation properties, direct application of arbitrary style patterns proves insufficient for reliable trigger implementation.

Our research addresses this limitation by introducing the \ourMethod{} framework for systematically selecting SET triggers.
We propose quantitative metrics, including Trigger Sensitivity and low-prevalence selection, to identify optimal style patterns, demonstrating that strategic trigger selection significantly enhances attack effectiveness compared to ad-hoc style application.
Furthermore, we explore synergistic effects of combining multiple style patterns as composite triggers and provide a comprehensive robustness analysis against style-normalization defenses.
Unlike previous approaches that focus primarily on identifier manipulation or dead code insertion, our work proposes a principled method for SET-based backdoor attacks, offering both theoretical foundations and practical implementation strategies.

\subsection{Backdoor Defenses}

Backdoor defenses can be broadly categorized into two primary approaches: \emph{data-level defenses} and \emph{model-level defenses}. While model-level defenses (which modify model architectures or training procedures to mitigate backdoor effects) have shown promise in image and text domains, they often require significant computational resources and may compromise model performance on clean inputs. In this work, we focus on data-level defenses, which aim to identify and remove poisoned samples from training data before model training, as they offer a more practical solution for code models. Data-level defenses typically fall into two categories: outlier-based methods and representation-based methods, each employing different strategies to distinguish between clean and poisoned samples.

\smallskip
\noindent\textbf{Outlier-based Methods.} Outlier-based methods identify poisoned samples by detecting statistical anomalies in the training data. Steinhardt et al.~\cite{steinhardt_certified_2017} and Paudice et al.~\cite{paudice_label_2018} developed detection techniques based on outlier removal, where they trained label-specific outlier detectors on trusted data to filter suspicious samples. Paudice et al.~\cite{paudice_detection_2018} proposed a different approach using k-nearest neighbors to relabel potentially poisoned samples to their most common neighboring class. More recent perturbation-based techniques have emerged, with Gao et al.~\cite{gao_strip_2019} introducing STRIP for image models, which analyzes prediction entropy under input perturbations to identify backdoored samples. For NLP models, Qi et al.~\cite{qi_onion_2021} developed ONION, leveraging language models to detect context-independent trigger words through a leave-one-out strategy.

\smallskip
\noindent\textbf{Representation-based Methods.} These methods exploit the latent representations of neural networks to distinguish between clean and poisoned samples. Chen et al.~\cite{chen_detecting_2019} introduced Activation Clustering, which applies K-means clustering to model activations to separate clean and poisoned samples based on their representation patterns.
Tran et al.~\cite{tran_spectral_2018} proposed Spectral Signatures, using singular value decomposition on activation patterns to identify contaminated data, which typically exhibits distinctive spectral characteristics.
Zhang et al.~\cite{zhang_backdoor_2023} developed a deconfounded representation learning approach that separates backdoor-related features from clean features to identify poisoned samples. Wang et al.~\cite{wang_training_2023} proposed a training set cleansing method using self-supervised representation learning to detect and filter out poisoned samples without relying on labeled data. Additionally, Zeng et al.~\cite{zhai_ncl_2023} introduced a contrastive learning framework that augments training data with noise to better distinguish between clean and poisoned examples in the representation space.

\smallskip
\noindent\textbf{Defenses for Code Models.}
Defense mechanisms specifically designed for neural code models remain relatively unexplored compared to CV and NLP domains. Existing approaches primarily adapt generic data-level defenses, such as activation clustering~\cite{chen_detecting_2019} and spectral signatures~\cite{tran_spectral_2018}, but these methods often demonstrate limited effectiveness in code domains~\cite{ramakrishnan_backdoors_2022,wan_you_2022,sun_backdooring_2023,yang_stealthy_2024}.
Recent advances include EliBadCode~\cite{sun_eliminating_2025} and KillBadCode~\cite{sun_show_2025}. EliBadCode is designed for identifier-based triggers, while KillBadCode operates by identifying anomalous tokens that disrupt code ``naturalness''. However, both defenses are ineffective against triggers that manifest as valid syntactic structures, such as non-identifier SET triggers (e.g., Style Types 2-9 in Table~\ref{tab:code_pattern}) and traditional dead-code insertions. EliBadCode's scope is strictly limited to identifiers. KillBadCode's n-gram-based naturalness model is susceptible to failure because these structural triggers are composed of locally plausible token sequences, even if the overarching structure is malicious or stylistically rare.
The primary limitation of existing defenses is their inability to effectively address diverse trigger types, particularly those exploiting structural and stylistic patterns beyond simple token-level modifications.

\subsection{Semantics-Preserving Code Transformations}

Semantics-preserving code transformations have been widely employed in various software engineering tasks, modifying code syntax or structure while preserving functionality. These transformations have proven valuable across multiple domains in software engineering research. Li et al.~\cite{li_ropgen_2022} utilized such transformations to enhance code classification models, demonstrating how structural variations improve model robustness while preserving code functionality. Bui et al.~\cite{bui_self-supervised_2021} leveraged code transformations for self-supervised learning of code representations, showing how semantically equivalent code variants help models learn more robust features. Rabin et al.~\cite{rabin_generalizability_2021} investigated how transformations affect the generalizability of code analysis tools, highlighting the importance of considering diverse code styles during model training. For data augmentation purposes, Wang et al.~\cite{wang_bridging_2022}, Li et al.~\cite{li_ropgen_2022}, and Yu et al.~\cite{yu_data_2022} employed various code transformations to increase training data diversity, demonstrating significant improvements in model performance through techniques such as variable renaming, statement reordering, and control flow restructuring. In security contexts, program obfuscation techniquesa form of semantics-preserving transformationhave been used to protect intellectual property but can also be exploited to evade malware detection~\cite{severi_explanation-guided_2021}.
Our work introduces a novel application of these transformations by systematically exploring their potential as backdoor triggers, representing a shift from viewing transformations as beneficial for model robustness to understanding their security implications.

\section{Conclusion}\label{sec:Conclusion}

The increasing adoption of neural code models in software engineering underscores the urgent need for awareness of emerging security threats.
This work introduces SET-based backdoor attacks, a novel kind of backdoor attacks that exploits imperceptible coding style variations as endogenous triggers, and presents \ourMethod{}, a principled framework for understanding their design, implementation, and characterization. In particular, we establish a principled method for SET-based backdoor attacks, grounded in five design principles and two quantitative criteria that enable systematic discovery and selection of effective triggers from the vast space of semantically-equivalent code transformations. Our systematic evaluation across neural code models, tasks, datasets, and programming languages demonstrates the effectiveness and stealth of SET-based attacks; especially, the high attack success rates (often exceeding 90\%) against the state-of-the-art defenses highlight the vulnerability of neural code models to SET-based backdoor attacks.
We showed that LLM-based code normalization can mitigate SET-based backdoor attacks to some extent. This calls for a community effort to investigate defenses that can completely thwart SET-based backdoor attacks.

\bibliographystyle{ACM-Reference-Format}
\bibliography{ref}

@String{Computing = "Computing" }

@String{Computer = "{IEEE} Computer" }

@String{Springer = "Springer-Verlag" }

@inproceedings{qi_onion_2021,
  author    = {Fanchao Qi and
               Yangyi Chen and
               Mukai Li and
               Yuan Yao and
               Zhiyuan Liu and
               Maosong Sun},
  title     = {{ONION:} {A} Simple and Effective Defense Against Textual Backdoor
               Attacks},
  booktitle = {Proceedings of the 2021 Conference on Empirical Methods in Natural
               Language Processing ({EMNLP}), Punta Cana, Dominican Republic},
  pages     = {9558--9566},
  publisher = {{ACL}},
  year      = {2021}
}

@inproceedings{zhou_devil_2023,
  author       = {Xin Zhou and
                  Kisub Kim and
                  Bowen Xu and
                  Jiakun Liu and
                  DongGyun Han and
                  David Lo},
  title        = {The Devil is in the Tails: How Long-Tailed Code Distributions Impact
                  Large Language Models},
  booktitle    = {38th {IEEE/ACM} International Conference on Automated Software Engineering,
                  {ASE} 2023, Luxembourg, September 11-15, 2023},
  pages        = {40--52},
  publisher    = {{IEEE}},
  year         = {2023}
}

@article{casalnuovo_studying_2019,
  author       = {Casey Casalnuovo and
                  Kenji Sagae and
                  Prem Devanbu},
  title        = {Studying the difference between natural and programming language corpora},
  journal      = {Empir. Softw. Eng.},
  volume       = {24},
  number       = {4},
  pages        = {1823--1868},
  year         = {2019}
}

@inproceedings{zhang_generating_2020,
  author       = {Huangzhao Zhang and
                  Zhuo Li and
                  Ge Li and
                  Lei Ma and
                  Yang Liu and
                  Zhi Jin},
  title        = {Generating Adversarial Examples for Holding Robustness of Source Code
                  Processing Models},
  booktitle    = {The Thirty-Fourth {AAAI} Conference on Artificial Intelligence, {AAAI}
                  2020, The Thirty-Second Innovative Applications of Artificial Intelligence
                  Conference, {IAAI} 2020, The Tenth {AAAI} Symposium on Educational
                  Advances in Artificial Intelligence, {EAAI} 2020, New York, NY, USA,
                  February 7-12, 2020},
  pages        = {1169--1176},
  publisher    = {{AAAI} Press},
  year         = {2020}
}

@inproceedings{du_extensive_2023,
  author       = {Xiaohu Du and
                  Ming Wen and
                  Zichao Wei and
                  Shangwen Wang and
                  Hai Jin},
  editor       = {Satish Chandra and
                  Kelly Blincoe and
                  Paolo Tonella},
  title        = {An Extensive Study on Adversarial Attack against Pre-trained Models
                  of Code},
  booktitle    = {Proceedings of the 31st {ACM} Joint European Software Engineering
                  Conference and Symposium on the Foundations of Software Engineering,
                  {ESEC/FSE} 2023, San Francisco, CA, USA, December 3-9, 2023},
  pages        = {489--501},
  publisher    = {{ACM}},
  year         = {2023}
}

@article{sun_eliminating_2025,
author = {Sun, Weisong and Chen, Yuchen and Fang, Chunrong and Feng, Yebo and Xiao, Yuan and Guo, An and Zhang, Quanjun and Chen, Zhenyu and Xu, Baowen and Liu, Yang},
title = {Eliminating Backdoors in Neural Code Models for Secure Code Understanding},
year = {2025},
issue_date = {July 2025},
publisher = {Association for Computing Machinery},
address = {New York, NY, USA},
volume = {2},
number = {FSE},
journal = {Proc. ACM Softw. Eng.},
month = jun,
articleno = {FSE063},
numpages = {23}
}

@article{Kocetkov2022TheStack,
  title={The Stack: 3 TB of permissively licensed source code},
  author={Kocetkov, Denis and Li, Raymond and Ben Allal, Loubna and Li, Jia and Mou,Chenghao and Muñoz Ferrandis, Carlos and Jernite, Yacine and Mitchell, Margaret and Hughes, Sean and Wolf, Thomas and Bahdanau, Dzmitry and von Werra, Leandro and de Vries, Harm},
  journal={Preprint},
  year={2022}
}

@misc{anton_starcoder_2024,
      title={StarCoder 2 and The Stack v2: The Next Generation}, 
      author={Anton Lozhkov and Raymond Li and Loubna Ben Allal and Federico Cassano and Joel Lamy-Poirier and Nouamane Tazi and Ao Tang and Dmytro Pykhtar and Jiawei Liu and Yuxiang Wei and Tianyang Liu and Max Tian and Denis Kocetkov and Arthur Zucker and Younes Belkada and Zijian Wang and Qian Liu and Dmitry Abulkhanov and Indraneil Paul and Zhuang Li and Wen-Ding Li and Megan Risdal and Jia Li and Jian Zhu and Terry Yue Zhuo and Evgenii Zheltonozhskii and Nii Osae Osae Dade and Wenhao Yu and Lucas Krauß and Naman Jain and Yixuan Su and Xuanli He and Manan Dey and Edoardo Abati and Yekun Chai and Niklas Muennighoff and Xiangru Tang and Muhtasham Oblokulov and Christopher Akiki and Marc Marone and Chenghao Mou and Mayank Mishra and Alex Gu and Binyuan Hui and Tri Dao and Armel Zebaze and Olivier Dehaene and Nicolas Patry and Canwen Xu and Julian McAuley and Han Hu and Torsten Scholak and Sebastien Paquet and Jennifer Robinson and Carolyn Jane Anderson and Nicolas Chapados and Mostofa Patwary and Nima Tajbakhsh and Yacine Jernite and Carlos Muñoz Ferrandis and Lingming Zhang and Sean Hughes and Thomas Wolf and Arjun Guha and Leandro von Werra and Harm de Vries},
      year={2024},
      eprint ={2402.19173},
      publisher = {arXiv}
}

@article{li_poison_2024,
  title = {Poison Attack and Poison Detection on Deep Source Code Processing Models},
  author = {Li, Jia and Li, Zhuo and Zhang, Huangzhao and Li, Ge and Jin, Zhi and Hu, Xing and Xia, Xin},
  year = {2024},
  journal = {ACM Trans. Softw. Eng. Methodol.},
  volume = {33},
  number = {3},
  pages = {62:1--62:31},
  issn = {1049-331X}
}

@inproceedings{zhai_ncl_2023,
  author       = {Shengfang Zhai and
                  Qingni Shen and
                  Xiaoyi Chen and
                  Weilong Wang and
                  Cong Li and
                  Yuejian Fang and
                  Zhonghai Wu},
  title        = {{NCL:} Textual Backdoor Defense Using Noise-Augmented Contrastive
                  Learning},
  booktitle    = {{IEEE} International Conference on Acoustics, Speech and Signal Processing
                  {ICASSP} 2023, Rhodes Island, Greece, June 4-10, 2023},
  pages        = {1--5},
  publisher    = {{IEEE}},
  year         = {2023}
}

@inproceedings{wang_training_2023,
  author       = {Hang Wang and
                  Sahar Karami and
                  Ousmane Dia and
                  Hippolyt Ritter and
                  Ehsan Emamjomeh{-}Zadeh and
                  Jiahui Chen and
                  Zhen Xiang and
                  David J. Miller and
                  George Kesidis},
  title        = {Training Set Cleansing of Backdoor Poisoning by Self-Supervised Representation
                  Learning},
  booktitle    = {{IEEE} International Conference on Acoustics, Speech and Signal Processing
                  {ICASSP} 2023, Rhodes Island, Greece, June 4-10, 2023},
  pages        = {1--5},
  publisher    = {{IEEE}},
  year         = {2023}
}

@inproceedings{zhang_backdoor_2023,
  title = {Backdoor Defense via Deconfounded Representation Learning},
  booktitle = {Proceedings of the {{IEEE}}/{{CVF Conference}} on {{Computer Vision}} and {{Pattern Recognition}},{{CVPR}}},
  author = {Zhang, Zaixi and Liu, Qi and Wang, Zhicai and Lu, Zepu and Hu, Qingyong},
  year = {2023},
  pages = {12228--12238}
}

@inproceedings{gao_strip_2019,
  author       = {Yansong Gao and
                  Chang Xu and
                  Derui Wang and
                  Shiping Chen and
                  Damith Chinthana Ranasinghe and
                  Surya Nepal},
  editor       = {David M. Balenson},
  title        = {{STRIP:} a defence against trojan attacks on deep neural networks},
  booktitle    = {Proceedings of the 35th Annual Computer Security Applications Conference,
                  {ACSAC} 2019, San Juan, PR, USA, December 09-13, 2019},
  pages        = {113--125},
  publisher    = {{ACM}},
  year         = {2019}
}

@inproceedings{paudice_label_2018,
  author       = {Andrea Paudice and
                  Luis Mu{\~{n}}oz{-}Gonz{\'{a}}lez and
                  Emil C. Lupu},
  editor       = {Carlos Alzate and
                  Anna Monreale and
                  Haytham Assem and
                  Albert Bifet and
                  Teodora Sandra Buda and
                  Bora Caglayan and
                  Brett Drury and
                  Eva Garc{\'{\i}}a{-}Mart{\'{\i}}n and
                  Ricard Gavald{\`{a}} and
                  Stefan Kramer and
                  Niklas Lavesson and
                  Michael Madden and
                  Ian M. Molloy and
                  Maria{-}Irina Nicolae and
                  Mathieu Sinn},
  title        = {Label Sanitization Against Label Flipping Poisoning Attacks},
  booktitle    = {{ECML} {PKDD} 2018 Workshops - Nemesis 2018, UrbReas 2018, SoGood
                  2018, IWAISe 2018, and Green Data Mining 2018, Dublin, Ireland, September
                  10-14, 2018, Proceedings},
  series       = {Lecture Notes in Computer Science},
  volume       = {11329},
  pages        = {5--15},
  publisher    = {Springer},
  year         = {2018}
}

@inproceedings{steinhardt_certified_2017,
  author       = {Jacob Steinhardt and
                  Pang Wei Koh and
                  Percy Liang},
  editor       = {Isabelle Guyon and
                  Ulrike von Luxburg and
                  Samy Bengio and
                  Hanna M. Wallach and
                  Rob Fergus and
                  S. V. N. Vishwanathan and
                  Roman Garnett},
  title        = {Certified Defenses for Data Poisoning Attacks},
  booktitle    = {Advances in Neural Information Processing Systems 30: Annual Conference
                  on Neural Information Processing Systems 2017, December 4-9, 2017,
                  Long Beach, CA, {USA}},
  pages        = {3517--3529},
  year         = {2017}
}

@article{paudice_detection_2018,
  author       = {Andrea Paudice and
                  Luis Mu{\~{n}}oz{-}Gonz{\'{a}}lez and
                  Andr{\'{a}}s Gy{\"{o}}rgy and
                  Emil C. Lupu},
  title        = {Detection of Adversarial Training Examples in Poisoning Attacks through
                  Anomaly Detection},
  journal      = {CoRR},
  volume       = {abs/1802.03041},
  year         = {2018},
  eprinttype    = {arXiv},
  eprint       = {1802.03041},
}

@inproceedings{svajlenko_2014_towards,
  title={Towards a big data curated benchmark of inter-project code clones},
  author={Svajlenko, Jeffrey and Islam, Judith F and Keivanloo, Iman and Roy, Chanchal K and Mia, Mohammad Mamun},
  booktitle={Proceedings of the 30th {IEEE} International Conference on Software Maintenance and Evolution (ICSME),
                  Victoria, BC, Canada},
  pages={476--480},
  year={2014}
}

@misc{he_45_2024,
  author    = {He, Hao and
               Yang, Haoqin and
               Burckhardt, Philipp and
               Kapravelos, Alexandros and
               Vasilescu, Bogdan and
               K{\"a}stner, Christian},
  title     = {4.5 Million (Suspected) Fake Stars in {GitHub}: A Growing Spiral of Popularity Contests, Scams, and Malware},
  year      = {2024},
  number    = {arXiv:2412.13459},
  eprint    = {2412.13459},
  archiveprefix = {arXiv}
}

@inproceedings{yi_badacts_2024,
  author       = {Yi, Biao and Chen, Sishuo and Li, Yiming and Li, Tong and
                  Zhang, Baolei and Liu, Zheli},
  title        = {{BadActs:} {A} Universal Backdoor Defense in the Activation Space},
  booktitle    = {Findings of the Association for Computational Linguistics, {ACL} 2024},
  pages        = {5339--5352},
  year         = {2024},
  address      = {Bangkok, Thailand},
  publisher    = {Association for Computational Linguistics}
}

@inproceedings{chen_expose_2022,
  author       = {Sishuo Chen and
                  Wenkai Yang and
                  Zhiyuan Zhang and
                  Xiaohan Bi and
                  Xu Sun},
  title        = {Expose Backdoors on the Way: {A} Feature-Based Efficient Defense against
                  Textual Backdoor Attacks},
  booktitle    = {Proceedings of Findings of the Association for Computational Linguistics: {EMNLP}
                  2022, Abu Dhabi, United Arab Emirates},
  pages        = {668--683},
  year         = {2022}
}

@article{gu_badnets_2019,
  title     = {{BadNets}: Identifying Vulnerabilities in the Machine Learning Model Supply Chain},
  author    = {Gu, Tianyu and {Dolan-Gavitt}, Brendan and Garg, Siddharth},
  journal={arXiv preprint arXiv:1708.06733},
  year      = {2019}
}

@misc{tree-sitter,
  author = {Brunsfeld, Max},
  title = {Tree-sitter},
  year = {2018},
  url = {https://tree-sitter.github.io/tree-sitter/}
}

@misc{copilot,
  title = {GitHub Copilot},
  year = {2023},
  url = {https://github.com/copilot/}
}

@misc{ijadata,
	author={{Ambient Software Evoluton Group}},
	title={IJaDataset 2.0 },
url= {http://secold.org/projects/seclone},
	month={Jan.},
	year={2013}
}

@inproceedings{RoziereLCL20Translation,
  author       = {Baptiste Rozi{\`{e}}re and
                  Marie{-}Anne Lachaux and
                  Lowik Chanussot and
                  Guillaume Lample},
  title        = {Unsupervised Translation of Programming Languages},
  booktitle    = {Proceedings of the Annual Conference
                  on Neural Information Processing Systems (NeuIPS), Virtual Event},
year      = {2020},
  pages     = {20601--20611}
}

@inproceedings{bui_self-supervised_2021,
  title     = {Self-Supervised Contrastive Learning for Code Retrieval and Summarization via Semantic-Preserving Transformations},
  booktitle = {Proceedings of the 44th International ACM SIGIR Conference on Research and Development in Information Retrieval, Virtual Event},
  author    = {Bui, Nghi D. Q. and Yu, Yijun and Jiang, Lingxiao},
  year      = {2021},
  pages     = {511--521}
}

@inproceedings{chen_badnl_2021,
  title     = {{BadNL}: Backdoor Attacks against NLP Models with Semantic-Preserving Improvements},
  booktitle = {Proceedings of the  37th Annual Computer Security Applications Conference (ACSAC), Virtual Event},
  author    = {Chen, Xiaoyi and Salem, Ahmed and Chen, Dingfan and Backes, Michael and Ma, Shiqing and Shen, Qingni and Wu, Zhonghai and Zhang, Yang},
  year      = {2021},
  pages     = {554--569}
}

@inproceedings{chen_detecting_2019,
  title     = {Detecting Backdoor Attacks on Deep Neural Networks by Activation Clustering},
  booktitle = {Proceedings of Workshop on Artificial Intelligence Safety 2019 Co-Located with the 33rd AAAI Conference on Artificial Intelligence 2019 (AAAI), Honolulu, Hawaii},
  author    = {Chen, Bryant and Carvalho, Wilka and Baracaldo, Nathalie and Ludwig, Heiko and Edwards, Benjamin and Lee, Taesung and Molloy, Ian M. and Srivastava, Biplav},
  year      = {2019}
}

@inproceedings{feng_codebert_2020,
  title     = {{CodeBERT}: A Pre-Trained Model for Programming and Natural Languages},
  booktitle = {Proceedings of Findings of the Association for Computational Linguistics: EMNLP, Virtual Event},
  author    = {Feng, Zhangyin and Guo, Daya and Tang, Duyu and Duan, Nan and Feng, Xiaocheng and Gong, Ming and Shou, Linjun and Qin, Bing and Liu, Ting and Jiang, Daxin and Zhou, Ming},
  year      = {2020},
  pages     = {1536--1547}
}

@misc{husain_codesearchnet_2020,
  title     = {CodeSearchNet Challenge: Evaluating the State of Semantic Code Search},
  author    = {Husain, Hamel and Wu, Ho-Hsiang and Gazit, Tiferet and Allamanis, Miltiadis and Brockschmidt, Marc},
  year      = {2020},
  number    = {arXiv:1909.09436},
  eprint    = {1909.09436},
  publisher = {arXiv}
}

@inproceedings{kurita_weight_2020,
  title     = {Weight Poisoning Attacks on Pretrained Models},
  booktitle = {Proceedings of the 58th Annual Meeting of the Association for Computational Linguistics (ACL), Virtual Event},
  author    = {Kurita, Keita and Michel, Paul and Neubig, Graham},
  year      = {2020},
  pages     = {2793--2806}
}

@article{li_invisible_2021,
  title   = {Invisible Backdoor Attacks on Deep Neural Networks via Steganography and Regularization},
  author  = {Li, Shaofeng and Xue, Minhui and Zhao, Benjamin Zi Hao and Zhu, Haojin and Zhang, Xinpeng},
  year    = {2021},
  journal = {IEEE Transactions on Dependable and Secure Computing},
  volume  = {18},
  number  = {5},
  pages   = {2088--2105},
  issn    = {1941-0018}
}

@inproceedings{li_multi-target_2023,
  title     = {Multi-Target Backdoor Attacks for Code Pre-Trained Models},
  booktitle = {Proceedings of the 61st Annual Meeting of the Association for Computational Linguistics (ACL), Toronto, Canada},
  author    = {Li, Yanzhou and Liu, Shangqing and Chen, Kangjie and Xie, Xiaofei and Zhang, Tianwei and Liu, Yang},
  year      = {2023},
  pages     = {7236--7254}
}

@inproceedings{li_ropgen_2022,
  title     = {{RoPGen}: Towards Robust Code Authorship Attribution via Automatic Coding Style Transformation},
  booktitle = {Proceedings of the 44th International Conference on Software Engineering (ICSE), Pittsburgh, PA, USA},
  author    = {Li, Zhen and Chen, Guenevere (Qian) and Chen, Chen and Zou, Yayi and Xu, Shouhuai},
  year      = {2022},
  pages     = {1906--1918}
}

@inproceedings{lin_composite_2020,
  title     = {Composite Backdoor Attack for Deep Neural Network by Mixing Existing Benign Features},
  booktitle = {Proceddings of the 27th ACM SIGSAC Conference on Computer and Communications Security (CCS), Virtual Event},
  author    = {Lin, Junyu and Xu, Lei and Liu, Yingqi and Zhang, Xiangyu},
  year      = {2020},
  pages     = {113--131}
}

@article{liu_combining_2021,
  title   = {Combining Graph Neural Networks with Expert Knowledge for Smart Contract Vulnerability Detection},
  author  = {Liu, Zhenguang and Qian, Peng and Wang, Xiaoyang and Zhuang, Yuan and Qiu, Lin and Wang, Xun},
  journal = {IEEE Transactions on Knowledge and Data Engineering},
  volume       = {35},
  number       = {2},
  pages        = {1296--1310},
  year         = {2023}
}

@inproceedings{xu_targeted_2021,
  author       = {Chang Xu and
                  Jun Wang and
                  Yuqing Tang and
                  Francisco Guzm{\'{a}}n and
                  Benjamin I. P. Rubinstein and
                  Trevor Cohn},
  editor       = {Jure Leskovec and
                  Marko Grobelnik and
                  Marc Najork and
                  Jie Tang and
                  Leila Zia},
  title        = {A Targeted Attack on Black-Box Neural Machine Translation with Parallel
                  Data Poisoning},
  booktitle    = {{WWW} '21: The Web Conference 2021, Virtual Event / Ljubljana, Slovenia,
                  April 19-23, 2021},
  pages        = {3638--3650},
  publisher    = {{ACM} / {IW3C2}},
  year         = {2021}
}

@inproceedings{barni_new_2019,
  author       = {Mauro Barni and
                  Kassem Kallas and
                  Benedetta Tondi},
  title        = {A New Backdoor Attack in {CNNS} by Training Set Corruption Without
                  Label Poisoning},
  booktitle    = {2019 {IEEE} International Conference on Image Processing, {ICIP} 2019,
                  Taipei, Taiwan, September 22-25, 2019},
  pages        = {101--105},
  publisher    = {{IEEE}},
  year         = {2019}
}

@inproceedings{liu_trojaning_2018,
  title     = {Trojaning Attack on Neural Networks},
  booktitle = {Proceedings of the 25th Annual Network and Distributed System Security Symposium (NDSS), San Diego, California, USA},
  author    = {Liu, Yingqi and Ma, Shiqing and Aafer, Yousra and Lee, Wen-Chuan and Zhai, Juan and Wang, Weihang and Zhang, Xiangyu},
  year      = {2018}
}

@inproceedings{lu_codexglue_2021,
  title   = {{CodeXGLUE}: A Machine Learning Benchmark Dataset for Code Understanding and Generation},
  author  = {Lu, Shuai and Guo, Daya and Ren, Shuo and Huang, Junjie and Svyatkovskiy, Alexey and Blanco, Ambrosio and Clement, Colin and Drain, Dawn and Jiang, Daxin and Tang, Duyu and Li, Ge and Zhou, Lidong and Shou, Linjun and Zhou, Long and Tufano, Michele and Gong, Ming and Zhou, Ming and Duan, Nan and Sundaresan, Neel and Deng, Shao Kun and Fu, Shengyu and Liu, Shujie},
  year    = {2021},
  booktitle = {Proceedings of the Neural Information Processing Systems Track on Datasets and Benchmarks, Virtual Event}
}

@inproceedings{gan_triggerless_2022,
  author       = {Leilei Gan and
                  Jiwei Li and
                  Tianwei Zhang and
                  Xiaoya Li and
                  Yuxian Meng and
                  Fei Wu and
                  Yi Yang and
                  Shangwei Guo and
                  Chun Fan},
  title        = {Triggerless Backdoor Attack for {NLP} Tasks with Clean Labels},
  booktitle    = {Proceedings of the 2022 Conference of the North American Chapter of
                  the Association for Computational Linguistics: Human Language Technologies,
                  {NAACL} 2022, Seattle, WA, United States, July 10-15, 2022},
  pages        = {2942--2952},
  publisher    = {Association for Computational Linguistics},
  year         = {2022}
}

@inproceedings{pan_hidden_2022,
  title     = {Hidden Trigger Backdoor Attack on NLP Models via Linguistic Style Manipulation},
  booktitle = {Proceedings of the 31st USENIX Security Symposium (USENIX Security), Boston,
                  MA, USA},
  author    = {Pan, Xudong and Zhang, Mi and Sheng, Beina and Zhu, Jiaming and Yang, Min},
  year      = {2022},
  pages     = {3611--3628}
}

@inproceedings{qi_hidden_2021,
  title     = {Hidden Killer: Invisible Textual Backdoor Attacks with Syntactic Trigger},
  booktitle = {Proceedings of the 59th Annual Meeting of the Association for Computational Linguistics and the 11th International Joint Conference on Natural Language Processing (Volume 1: Long Papers), Virtual Event},
  author    = {Qi, Fanchao and Li, Mukai and Chen, Yangyi and Zhang, Zhengyan and Liu, Zhiyuan and Wang, Yasheng and Sun, Maosong},
  year      = {2021},
  pages     = {443--453}
}

@inproceedings{qi_turn_2021,
  title     = {Turn the Combination Lock: Learnable Textual Backdoor Attacks via Word Substitution},
  booktitle = {Proceedings of the 59th Annual Meeting of the Association for Computational Linguistics and the 11th International Joint Conference on Natural Language Processing (Volume 1: Long Papers), Virtual Event},
  author    = {Qi, Fanchao and Yao, Yuan and Xu, Sophia and Liu, Zhiyuan and Sun, Maosong},
  year      = {2021},
  pages     = {4873--4883}
}

@article{rabin_generalizability_2021,
  title   = {On the Generalizability of Neural Program Models with Respect to Semantic-Preserving Program Transformations},
  author  = {Rabin, Md Rafiqul Islam and Bui, Nghi D. Q. and Wang, Ke and Yu, Yijun and Jiang, Lingxiao and Alipour, Mohammad Amin},
  year    = {2021},
  journal = {Information and Software Technology},
  volume  = {135},
  pages   = {106552},
  issn    = {0950-5849}
}

@inproceedings{ramakrishnan_backdoors_2022,
  title     = {Backdoors in Neural Models of Source Code},
  booktitle = {Proceedings of the 26th International Conference on Pattern Recognition (ICPR), Montreal, QC, Canada},
  author    = {Ramakrishnan, Goutham and Albarghouthi, Aws},
  year      = {2022},
  pages     = {2892--2899}
}

@inproceedings{aghakhani_trojanPuzzle_2024,
  author       = {Hojjat Aghakhani and
                  Wei Dai and
                  Andre Manoel and
                  Xavier Fernandes and
                  Anant Kharkar and
                  Christopher Kruegel and
                  Giovanni Vigna and
                  David Evans and
                  Ben Zorn and
                  Robert Sim},
  title        = {TrojanPuzzle: Covertly Poisoning Code-Suggestion Models},
  booktitle    = {{IEEE} Symposium on Security and Privacy, {SP} 2024, San Francisco,
                  CA, USA, May 19-23, 2024},
  pages        = {1122--1140},
  publisher    = {{IEEE}},
  year         = {2024}
}

@inproceedings{schuster_you_2021,
  title     = {You Autocomplete Me: Poisoning Vulnerabilities in Neural Code Completion},
  booktitle = {Proceedings of the 30th USENIX Security Symposium (USENIX Security), Virtual Event},
  author    = {Schuster, Roei and Song, Congzheng and Tromer, Eran and Shmatikov, Vitaly},
  year      = {2021},
  pages     = {1559--1575}
}

@inproceedings{severi_explanation-guided_2021,
  title     = {Explanation-Guided Backdoor Poisoning Attacks against Malware Classifiers},
  booktitle = {Proceddings of the 30th USENIX Security Symposium (USENIX Security), Virtual Event},
  author    = {Severi, Giorgio and Meyer, Jim and Coull, Scott and Oprea, Alina},
  year      = {2021},
  pages     = {1487--1504}
}

@inproceedings{sun_backdooring_2023,
  title     = {Backdooring Neural Code Search},
  booktitle = {Proceedings of the 61st Annual Meeting of the Association for Computational Linguistics (ACL), Toronto, Canada},
  author    = {Sun, Weisong and Chen, Yuchen and Tao, Guanhong and Fang, Chunrong and Zhang, Xiangyu and Zhang, Quanjun and Luo, Bin},
  year      = {2023},
  pages     = {9692--9708}
}

@inproceedings{sun_code_2022,
  title     = {Code Search Based on Context-Aware Code Translation},
  booktitle = {Proceedings of the 44th International Conference on Software Engineering (ICSE), Pittsburgh, PA, USA},
  author    = {Sun, Weisong and Fang, Chunrong and Chen, Yuchen and Tao, Guanhong and Han, Tingxu and Zhang, Quanjun},
  year      = {2022},
  pages     = {388--400}
}

@inproceedings{sun_codemark_2023,
  title     = {CodeMark: Imperceptible Watermarking for Code Datasets against Neural Code Completion Models},
  booktitle = {Proceedings of the 31st ACM Joint European Software Engineering Conference and Symposium on the Foundations of Software Engineering},
  author    = {Sun, Zhensu and Du, Xiaoning and Song, Fu and Li, Li},
  year      = {2023},
  eprint    = {2308.14401},
  pages     = {1561--1572}
}

@inproceedings{tran_spectral_2018,
  title     = {Spectral Signatures in Backdoor Attacks},
  booktitle = {Proceedings of the Advances in Neural Information Processing Systems (NeurIPS), Montr{\'{e}}al, Canada},
  author    = {Tran, Brandon and Li, Jerry and Madry, Aleksander},
pages        = {8011--8021},
  year      = {2018}
}

@article{tufano_empirical_2019,
  title     = {An Empirical Study on Learning Bug-Fixing Patches in the Wild via Neural Machine Translation},
  author    = {Tufano, Michele and Watson, Cody and Bavota, Gabriele and Penta, Massimiliano Di and White, Martin and Poshyvanyk, Denys},
  year      = {2019},
  journal   = {ACM Transactions on Software Engineering and Methodology},
  volume    = {28},
  number    = {4},
  pages     = {19:1--19:29}
}

@inproceedings{wan_you_2022,
  title     = {You See What {I} Want You to See: Poisoning Vulnerabilities in Neural Code Search},
  booktitle = {Proceedings of the 30th ACM Joint European Software Engineering Conference and Symposium on the Foundations of Software Engineering (ESEC/FSE), Singapore},
  author    = {Wan, Yao and Zhang, Shijie and Zhang, Hongyu and Sui, Yulei and Xu, Guandong and Yao, Dezhong and Jin, Hai and Sun, Lichao},
  year      = {2022},
  pages     = {1233--1245}
}

@article{yang_stealthy_2024,
  author={Yang, Zhou and Xu, Bowen and Zhang, Jie M. and Kang, Hong Jin and Shi, Jieke and He, Junda and Lo, David},
  journal={IEEE Transactions on Software Engineering}, 
  title={Stealthy Backdoor Attack for Code Models}, 
  year={2024},
  volume={50},
  number={4},
  pages={721-741}
}

@inproceedings{wang_bridging_2022,
  title     = {Bridging Pre-Trained Models and Downstream Tasks for Source Code Understanding},
  booktitle = {Proceedings of the 44th International Conference on Software Engineering (ICSE), Pittsburgh, PA, USA},
  author    = {Wang, Deze and Jia, Zhouyang and Li, Shanshan and Yu, Yue and Xiong, Yun and Dong, Wei and Liao, Xiangke},
  year      = {2022},
  pages     = {287--298}
}

@inproceedings{wang_codet5_2021,
  title     = {{CodeT5}: Identifier-Aware Unified Pre-Trained Encoder-Decoder Models for Code Understanding and Generation},
  booktitle = {Proceedings of the 2021 Conference on Empirical Methods in Natural Language Processing, Punta Cana (EMNLP), Dominican
                  Republic},
  author    = {Wang, Yue and Wang, Weishi and Joty, Shafiq and Hoi, Steven C.H.},
  year      = {2021},
  pages     = {8696--8708}
}

@inproceedings{wang_invisible_2022,
  title     = {An Invisible Black-Box Backdoor Attack through Frequency Domain},
  booktitle = {Proceedings of the 17th European Conference Computer Vision (ECCV), Tel Aviv,
                  Israel},
  author    = {Wang, Tong and Yao, Yuan and Xu, Feng and An, Shengwei and Tong, Hanghang and Wang, Ting},
  year      = {2022},
  pages     = {396--413}
}

@article{yu_data_2022,
  title   = {Data Augmentation by Program Transformation},
  author  = {Yu, Shiwen and Wang, Ting and Wang, Ji},
  year    = {2022},
  journal = {Journal of Systems and Software},
  volume  = {190},
  pages   = {111304},
  issn    = {0164-1212}
}

@article{zhang_poison_2022,
  title   = {{Poison Ink}: Robust and Invisible Backdoor Attack},
  author  = {Zhang, Jie and Dongdong, Chen and Huang, Qidong and Liao, Jing and Zhang, Weiming and Feng, Huamin and Hua, Gang and Yu, Nenghai},
  year    = {2022},
  journal = {IEEE Transactions on Image Processing},
  volume  = {31},
  pages   = {5691--5705},
  issn    = {1057-7149, 1941-0042}
}

@inproceedings{zhou_devign_2019,
  title     = {Devign: Effective Vulnerability Identification by Learning Comprehensive Program Semantics via Graph Neural Networks},
  booktitle = {Proceedings of the Annual Conference on Neural Information Processing Systems (NeurIPS), Vancouver, BC, Canada},
  author    = {Zhou, Yaqin and Liu, Shangqing and Siow, Jing Kai and Du, Xiaoning and Liu, Yang},
  year      = {2019},
  pages     = {10197--10207}
}

@article{liu_backdoor_2025,
  title={Backdoor threats in large language models—a survey},
  author={Liu, Shuai and Pan, Yiheng and Hong, Kun and Fei, Ruite and Lin, Chenhao and Li, Qian and Shen, Chao},
  journal={Science China Information Sciences},
  volume={68},
  number={9},
  pages={1--34},
  year={2025},
  publisher={Springer}
}

@article{chen_deep_2025,
  title={Deep learning-based software engineering: progress, challenges, and opportunities},
  author={Chen, Xiangping and Hu, Xing and Huang, Yuan and Jiang, He and Ji, Weixing and Jiang, Yanjie and Jiang, Yanyan and Liu, Bo and Liu, Hui and Li, Xiaochen and others},
  journal={Science China Information Sciences},
  volume={68},
  number={1},
  pages={111102},
  year={2025},
  publisher={Springer}
}

@inproceedings{sun_show_2025,
  author       = {Weisong Sun and
                  Yuchen Chen and
                  Mengzhe Yuan and
                  Chunrong Fang and
                  Zhenpeng Chen and
                  Chong Wang and
                  Yang Liu and
                  Baowen Xu and
                  Zhenyu Chen},
  title        = {Show Me Your Code! Kill Code Poisoning: {A} Lightweight Method Based
                  on Code Naturalness},
  booktitle    = {47th {IEEE/ACM} International Conference on Software Engineering,
                  {ICSE} 2025, Ottawa, ON, Canada, April 26 - May 6, 2025},
  pages        = {2663--2675},
  publisher    = {{IEEE}},
  year         = {2025}
}

@inproceedings{hindle_naturalness_2012,
  author       = {Abram Hindle and
                  Earl T. Barr and
                  Zhendong Su and
                  Mark Gabel and
                  Premkumar T. Devanbu},
  editor       = {Martin Glinz and
                  Gail C. Murphy and
                  Mauro Pezz{\`{e}}},
  title        = {On the naturalness of software},
  booktitle    = {34th International Conference on Software Engineering, {ICSE} 2012,
                  June 2-9, 2012, Zurich, Switzerland},
  pages        = {837--847},
  publisher    = {{IEEE} Computer Society},
  year         = {2012}
}

@inproceedings{yefet_adversarial_2020,
  author       = {Noam Yefet and
                  Uri Alon and
                  Eran Yahav},
  title        = {Adversarial examples for models of code},
  journal      = {Proc. {ACM} Program. Lang.},
  volume       = {4},
  number       = {{OOPSLA}},
  pages        = {162:1--162:30},
  year         = {2020}
}

@inproceedings{buse_learning_2010,
  author       = {Raymond P. L. Buse and
                  Westley Weimer},
  title        = {Learning a Metric for Code Readability},
  journal      = {{IEEE} Trans. Software Eng.},
  volume       = {36},
  number       = {4},
  pages        = {546--558},
  year         = {2010}
}

\typeout{get arXiv to do 4 passes: Label(s) may have changed. Rerun}
\end{document}